\documentclass{aa}

\usepackage{diagbox,hyperref,makecell,txfonts}
\hypersetup{colorlinks,citecolor=blue,linkcolor=blue,urlcolor=blue}
\graphicspath{{Figures/}}

\defcitealias{EHT1}{EHT L1}
\defcitealias{EHT2}{EHT L2}
\defcitealias{EHT3}{EHT L3}
\defcitealias{EHT4}{EHT L4}
\defcitealias{EHT5}{EHT L5}
\defcitealias{EHT6}{EHT L6}
\defcitealias{EHT7}{EHT L7}
\defcitealias{EHT8}{EHT L8}
\defcitealias{Gralla2020}{GLM 20}

\newcommand{\ab}[1]{\left|#1\right|}
\newcommand{\av}[1]{\left\langle#1\right\rangle}
\newcommand{\br}[1]{\left[#1\right]}
\newcommand{\pa}[1]{\left(#1\right)}
\newcommand{\dt}{\mathop{}\!\delta}
\newcommand{\ed}{\mathop{}\!\mathrm{d}}
\newcommand{\pd}{\mathop{}\!\boldsymbol{\partial}}\DeclareMathOperator\sign{sign}
\DeclareMathOperator\sn{sn}

\begin{document}

\title{Images and photon ring signatures\\
of thick disks around black holes}

\author{ 
    Frederic H. Vincent\inst{1}
    \and
    Samuel E. Gralla\inst{2}
    \and
    Alexandru Lupsasca\inst{3}
    \and
    Maciek Wielgus\inst{4}
}

\institute{
    LESIA, Observatoire de Paris, Universit\'e PSL, CNRS, Sorbonne Universit\'es, UPMC Univ. Paris 06, Univ. de Paris, Sorbonne Paris Cit\'e, 5 place Jules Janssen, 92195 Meudon, France.
    \email{frederic.vincent@obspm.fr}
    \and
    Department of Physics, University of Arizona, Tucson, AZ 85721, USA
    \and
    Princeton Gravity Initiative, Princeton University, Princeton, NJ 08544, USA
    \and
    Max-Planck-Institut f\"ur Radioastronomie, Auf dem H\"ugel 69, D-53121 Bonn, Germany
}

\abstract
{High-frequency very-long-baseline interferometry (VLBI) observations can now resolve the event-horizon-scale emission from sources in the immediate vicinity of nearby supermassive black holes.
Future space-VLBI observations will access highly lensed features of black hole images---\textit{photon rings}---that will provide particularly sharp probes of strong-field gravity.}
{Focusing on the particular case of the supermassive black hole M87*, our goal is to explore a wide variety of accretion flows onto a Kerr black hole and to understand their corresponding images and visibilities.
We are particularly interested in the visibility on baselines to space, which encodes the photon ring shape and whose measurement could provide a stringent test of the Kerr hypothesis.}
{We develop a fully analytical model of stationary, axisymmetric accretion flows with a variable disk thickness and a matter four-velocity that can smoothly interpolate between purely azimuthal rotation and purely radial infall.
To determine the observational appearance of such flows, we numerically integrate the general-relativistic radiative transfer equation in the Kerr spacetime, taking care to include the effects of thermal synchrotron emission and absorption.
We then Fourier transform the resulting images and analyze their visibility amplitudes along the directions parallel and orthogonal to the black hole spin projected on the observer sky.}
{Our images generically display a ``wedding cake'' structure composed of discrete, narrow photon rings $(n=1,2,\ldots)$ stacked on top of broader primary emission that surrounds a central brightness depression of model-dependent size.
At 230\,GHz, the $n=1$ ring is always visible, but the $n=2$ ring is sometimes suppressed due to absorption.
At 345\,GHz, the medium is optically thinner and the $n=2$ ring displays clear signatures in both the image and visibility domains.
We also examine the thermal synchrotron emissivity in the equatorial plane and show that it exhibits an exponential dependence on radius for the preferred M87* parameters.}
{The ``black hole shadow'' is a model-dependent phenomenon---even for diffuse, optically thin sources---and should not be regarded as a generic prediction of general relativity.
Observations at 345\,GHz are promising for future space-VLBI measurements of the photon ring shape, since at this frequency the signal of the $n=2$ ring persists despite the disk thickness and non-zero absorption featured in our models.
Future work is needed to investigate whether this conclusion holds in a larger variety of reasonable models.}


\maketitle

\section{Introduction}

In 2019, the Event Horizon Telescope collaboration released 1.3\,mm interferometric observations of the supermassive black hole M87* at the center of the galaxy Messier 87 \citepalias[EHT;][]{EHT1}, achieving an effective angular resolution comparable to the black hole size.
These observations revealed the presence of a bright ring of approximately $40\,\mu$as in diameter that surrounds a much darker central region.
While these basic image features are undisputed and have in fact been independently confirmed \citep{Arras2022,Carilli2022,Lockhart2022}, several aspects of their theoretical interpretation remain open.
For example, should the dark area be associated with the ``black hole shadow'' \citep{Falcke2000}, as originally proposed \citepalias{EHT1}, or with the apparent position of the equatorial event horizon, as the data now seem to suggest \citep{Chael2021}?

\begin{figure*}
	\centering
	\includegraphics[width=0.95\textwidth]{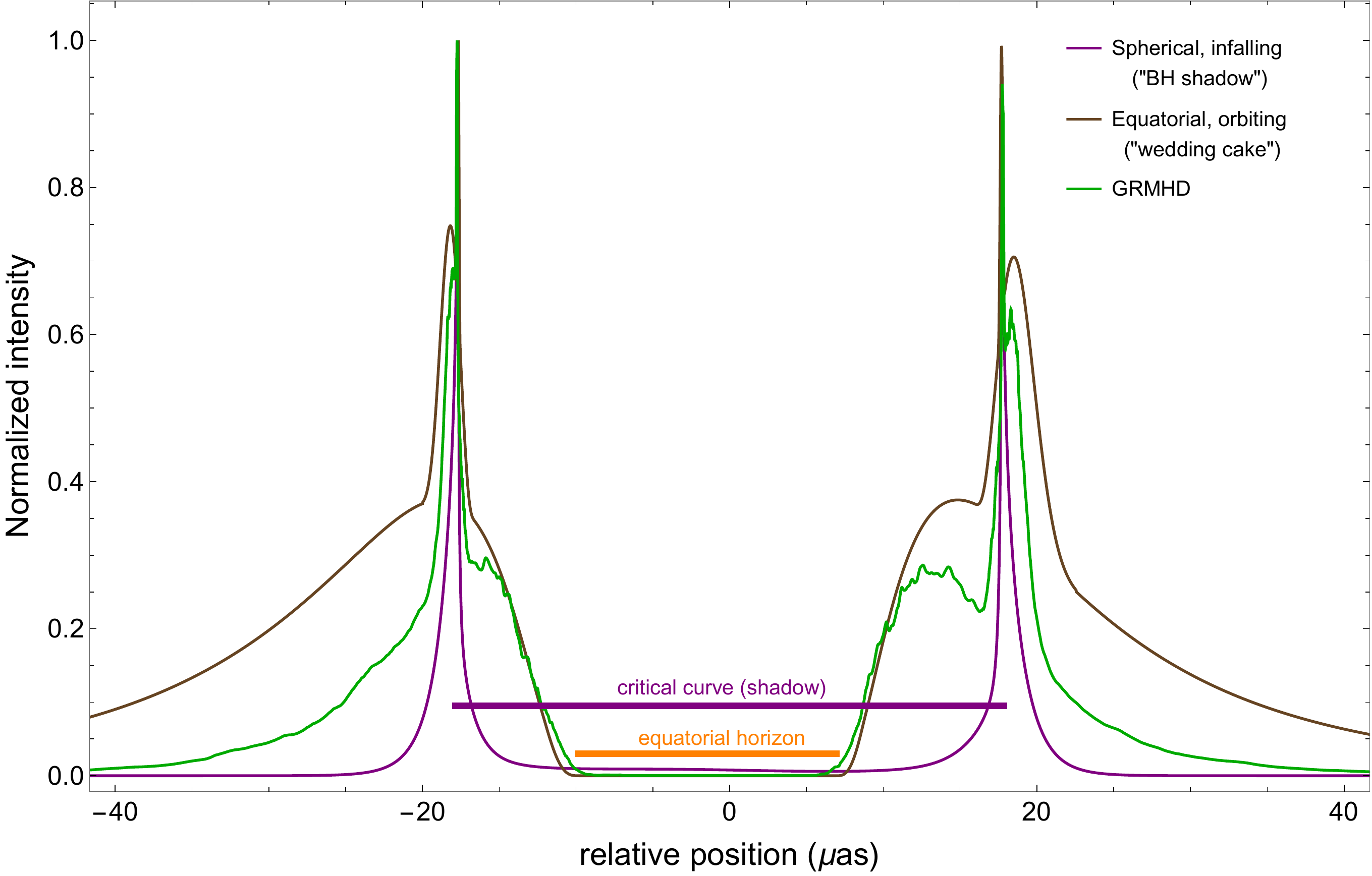} 
	\caption{Under the assumption of optically thin emitting matter concentrated very near the horizon, the range of reasonable appearances for models of accretion onto a Kerr black hole can be bracketed by two extreme toy models: equatorial, orbiting matter (brown), which produces a ``wedding cake'' structure \citep{Gralla2019,Johnson2019}, and spherical, infalling matter (purple), which produces a ``shadow'' \citep{Falcke2000}.
	Here, we show a 230\,GHz intensity cut parallel to the spin axis of M87* (taken to be a rapidly spinning black hole with $a=0.94$), including photons that orbit at most one full orbit around the black hole (up to $n=2$ half-orbits), with each model normalized to its peak intensity.
	The brown and purple curves were ray traced from analytic models (see Sec.~\ref{subsec:IntensityCuts} for details), while the green curve is numerical data from \citet{Johnson2019} produced with the simulation pipeline described in \citet{Wong2022}.
	The horizontal bars indicate the location of the Kerr critical curve and the apparent position of the equatorial event horizon.} 
	\label{fig:Menu}
\end{figure*}

A plausible range of observational appearances for M87* is depicted in Fig.~\ref{fig:Menu}.
Under the currently favored assumption of optically thin emission concentrated very near the event horizon \citepalias[e.g.,][]{EHT5,EHT8}, the characteristic appearance ranges between two extremes: a narrow photon ring surrounding a dark region inside the critical curve [a ``shadow''---see \citet{Falcke2000}], and a series of discrete photons rings stacked on top of broader emission outside the apparent equatorial horizon [a ``wedding cake''---see \citet[][Fig.~1]{Gralla2019} and \citet[][Fig.~3]{Johnson2019}].
Models of spherically symmetric, infalling matter lead to shadows, while models with equatorial, orbiting matter generate wedding cakes.
General-relativistic magnetohydrodynamic (GRMHD) models generally favor the wedding cake over the shadow \citep{Johnson2019,Chael2021}, but the debate has yet to be settled \citep{Bronzwaer2021}.

The purpose of this paper is to more fully flesh out the model space between the shadow and wedding cake extremes.
Having a broad range of models is crucial not just for understanding the source but also for accurately forecasting its potential for future observations.
We are especially motivated by the promise of photon ring (orbiting light) measurements on long interferometric baselines \citep{Johnson2019,Gralla2020b}, which could provide a stringent test of the Kerr hypothesis \citep{Gralla2020,Wielgus2021}.
A recent proposal to measure the precise shape of the $n=2$ photon ring (formed by light that executes a full orbit around the black hole before escaping to a detector) relied exclusively on a purely equatorial and perfectly absorption-free model of the emission (\citet{Gralla2020}---henceforth \citetalias{Gralla2020}).
It is imperative to check whether this exciting prospect survives the introduction of geometrical thickness, absorption, and other potential complications of the real astrophysical flow \citep[see also][]{Paugnat2022}.

In this paper, we construct a set of semi-analytic models that bridge the gap between the shadow and wedding cake extremes, while also including a ``realistic'' level of absorption.
For select density, temperature, and velocity profiles, we compute the synchrotron emission and absorption based on the assumption of a uniformly magnetized disk.
Tuning parameters to match the total 230\,GHz horizon-scale flux density reported by EHT, we perform radiative transport to determine the observational appearance at both 230\,GHz and 345\,GHz, the planned frequency of future observations \citepalias[\citealt{ngEHT}; ][]{EHT2}.
Finally, we Fourier transform the images and check whether the photon ring signatures are observable on moderate and long baselines.
We consider both infalling and orbiting matter, as well as geometrically thin, thick, and fully spherical emission regions.
We fix the observer inclination to a value of $160^\circ$, deemed to be very likely for M87* \citep{Walker2018}, while varying the black hole spin from near-zero to near-maximal spin.

Our results have both theoretical and practical implications.
On the theory side, we confirm that the size of the central brightness depression is highly model-dependent, while the presence of discrete photon rings is generic \citep{Gralla2019,Johnson2019}.
We also confirm that the classic ``black hole shadow'' (a sharp intensity drop inside the critical curve) is caused by extreme special-relativistic redshifting \citep{Narayan2019,Gralla2021}, depending only indirectly on general-relativistic effects.
Furthermore, we find that the intensity drop occurs only in fully spherical, infalling models, whereas the dark area is highly distinct from the critical curve even for very thick disks assumed to contain purely infalling matter.
That is, far from being a generic prediction for optically thin flows \citep[as claimed, e.g., in][]{Psaltis2019,Narayan2019}, the appearance of a dark ``shadow'' filling the interior of the critical curve is in fact a \textit{remarkably fine-tuned} phenomenon.
This provides further evidence that the EHT observations should not be regarded as the black hole shadow in the strict sense given by \citet{Falcke2000}.

On the practical side, our results indicate that the $n=2$ photon ring is only marginally observable at 230\,GHz.
For strongly magnetized disks, we reproduce previous results \citep{Johnson2019,Chael2021} showing a prominent $n=2$ photon ring, but we find that in the weak-magnetization regime, the strength of the $n=2$ signal is sensitive to the astrophysical details of the source, and especially to the black hole spin.
In some cases, the $n=2$ signal even vanishes entirely due to absorption.
However, we find that at 345\,GHz, the $n=2$ signal returns to levels broadly consistent with previous estimates.
This suggests that 345\,GHz is an appropriate target for future space-VLBI observations of M87*, since the signal at that frequency is robust to the astrophysical parameters we vary.
We stress that these results only mark the beginning of a systematic study of the reasonable parameter space.
In particular, our models exclude the possibilities of tilted disks, highly inclined observers, or jet-base emission.

The paper is organized as follows.
We present our accretion disk model in Sec.~\ref{sec:Model} and display images of some accretion flows in Sec.~\ref{sec:Images}, before discussing the resulting visibility amplitudes in Sec.~\ref{sec:Visibilities} and summarizing our conclusions in Sec.~\ref{sec:Conclusion}.

\section{Accretion flow model}
\label{sec:Model}

In this paper, we consider accretion onto a black hole described by the Kerr metric in Boyer-Lindquist coordinates $(t,r,\theta,\phi)$.
The dimensionless spin parameter is labeled $a$.
We use natural units such that $G=c=1$, and radii are thus expressed in units of the black hole mass $M$.
We always use $\rho$ to denote the cylindrical radius $\rho=r\sin{\theta}$ (rather than a density) and $z=r\cos{\theta}$ to denote the cylindrical height above the equatorial plane $\theta=\pi/2$.  

\subsection{Disk geometry, physical quantities, and emission profile}

Our goal is to develop a very general and fully analytical model of geometrically thick disks, extending the one in \citet{Vincent2021}.
We restrict our attention to axisymmetric disks, so there is no dependence on $\phi$ in any of our physical quantities.

We consider a population of thermal electrons that fills the spacetime outside the event horizon.
The (fluid-frame) electron number density $n_\mathrm{e}(\rho,z)$ is specified in the equatorial plane at the (cylindrical) horizon radius $\rho_\mathrm{H}=r_\mathrm{H}=1+\sqrt{1-a^2}$,
\begin{align}
    n_\mathrm{e;H}=n_\mathrm{e}(\rho_\mathrm{H},z=0).
\end{align}
as is the electron temperature $T_\mathrm{e}(\rho,z)$,
\begin{align}
    T_\mathrm{e;H}=T_\mathrm{e}(\rho_\mathrm{H},z=0).
\end{align}
We define density and temperature profiles via a prescription similar to the one used in the analytical radiatively inefficient accretion flow (RIAF) models \citep[see, e.g.,][]{Broderick2011}:
\begin{subequations}
\label{eq:Prescription}
\begin{align}
    n_\mathrm{e}(r,z)&=n_\mathrm{e;H}\pa{\frac{r}{r_\mathrm{H}}}^{-2}\exp\pa{-\frac{z^2}{2(\alpha\rho)^2}},\\
    T_\mathrm{e}(r)&=T_\mathrm{e;H}\pa{\frac{r}{r_\mathrm{H}}}^{-1}.
\end{align}
\end{subequations}
The dependence of the density profile on the cylindrical height $z$ is thus a simple Gaussian with standard deviation $s_z=\alpha\rho$, where $\alpha$ is a free parameter that sets the opening angle of the accretion disk.
Indeed, if we define the surface of the disk to be a cone of height $s_z$ above the equatorial plane, then $\alpha$ corresponds to the tangent of its opening angle.
As $\alpha\to0$, we recover equatorial models similar to those of \citetalias{Gralla2020} and \citet{Paugnat2022}, but here we keep $\alpha\neq0$ to explore the effects of disk thickness, which is expected to be significant for realistic RIAF flows \citep{Yuan2014}.

The magnitude of the magnetic field, which is necessary to compute the synchrotron radiation, is prescribed by imposing a constant magnetization throughout the disk.
That is, we fix the ratio of the magnetic field and particle energy densities
\begin{align}
    \label{eq:Magnetization}
    \sigma=\frac{B^2/4\pi}{m_\mathrm{p}c^2n_\mathrm{e}}.
\end{align}
where $B$ is the magnetic field magnitude, $m_\mathrm{p}$ the proton mass, and $c$ the speed of light (which we have restored here for clarity).

We include the effects of thermal synchrotron emission and absorption using the phenomenological expression derived by \citet{Leung2011}.
We present this expression and discuss its accuracy in App.~\ref{app:Synchrotron}.
As shown in App.~\ref{app:Emissivity}, the radial profile of thermal synchrotron emission in our model is approximately
\begin{align}
    \label{eq:EmissionProfile}
    j_\nu(r)\propto\mathrm{exp}\pa{-\zeta\frac{r}{r_\mathrm{H}}},
\end{align}
where $\zeta$ is a model-dependent number that takes a typical value near 3 for our models.
However, we emphasize that Eq.~\eqref{eq:EmissionProfile} is used for interpretation only; our computations are performed with the more precise expression \eqref{eq:HighTemperatureEmissivity} of \citet{Leung2011}.

\subsection{Dynamics}

\subsubsection{Orbiting motion}
\label{subsec:CircularMotion}

To compute images, we still need to specify the four-velocity of the emitting electrons in the accretion flow.
We introduce a linear combination of an orbiting matter component and an infalling matter component that allows one to interpolate between the two extremes illustrated in Fig.~\ref{fig:Menu}.

First, we define the orbiting component by an azimutal four-velocity field with vanishing radial and polar components.
As discussed in App.~\ref{app:CircularMotion}, it is nontrivial to find a prescription that is simple, everywhere smooth and becomes Keplerian at large distances from the black hole.
We follow the prescription of~\citet{Gold2020} and write the four-velocity 1-form as
\begin{align}
    \label{eq:CircularFourVelocity}
    u^\mathrm{circ}_\mu\ed x^\mu&=-u^\mathrm{circ}_t\pa{-\ed t+\ell\ed\phi},\qquad
    \ell=\frac{\rho^{3/2}}{1+\rho},
\end{align}
which is a special case of the general profile \eqref{eq:lalpha} described in App.~\ref{app:CircularMotion}.
This choice gives rise to a mild divergence $\Omega\sim\rho^{-1/2}$ at the poles, but is otherwise well-defined outside the horizon.
The polar divergence is irrelevant for our models of thin and thick disks, as they have effectively no emission from the poles, and it has no noticeable effect even in the limit of fully spherical emission for infalling matter (Fig.~\ref{fig:WeddingCakes} left). 

Unit-normalization of the four-velocity \eqref{eq:CircularFourVelocity} requires
\begin{align}
	-u^\mathrm{circ}_t=\frac{1}{\sqrt{-\pa{g^{tt}-2g^{t\phi}\ell+g^{\phi\phi}\ell^2}}}.
\end{align}
Note that this circular motion is not geodesic.

\renewcommand{\arraystretch}{1.3}
\begin{table*}[h!]
    \centering
    \begin{tabular}{|c|c|c|} 
    \hline
    Symbol &  Value & Property \\
    \hline
    $M$ & $6.2\times10^9\,M_\odot$ & compact object mass \\
    $D$ & 16.9\,Mpc & compact object distance \\
    $a$ & $\{0.01,0.94\}$ & BH spin parameter \\
    $\alpha=\tan{\theta_\mathrm{op}}$ & $\{0.1,1\}$ & disk opening angle \\
    $n_\mathrm{e;H}$ &
    \begin{tabular}{|cc|}
	    \hline
        circ/thin & $\{1.5,5\}$ \\
        \hline
        rad/thin & $\{5,20\}$ \\
        \hline
        circ/thick & $\{0.7,2\}$ \\
        \hline
        rad/thick & $\{2,10\}$ \\
        \hline
    \end{tabular} & max density of electrons \\
    $T_\mathrm{e;H}$ & $10^{11}$\,K & max electron temperature \\
    $\sigma$ & 0.01 & magnetization \\
    $i$ & $160^\circ$ & inclination angle \\
    $\nu_\mathrm{obs}$ & $[230,345]$\,GHz & observing frequency \\
    $f$ & $100\,\mu$as & field of view \\
    $N\times N$ & $20000\times20000$ & image resolution \\
    \hline
    \end{tabular}
    \caption{Parameters of our model.
    The maximum electron number density varies across models in order to ensure an observed flux of $\approx 0.5$\,Jy at 230\,GHz.
    It is expressed in units of $10^6$\,cm$^{-3}$, with the first and second numbers referring to BH spins of $a=0.01$ and $a=0.94$, respectively.
    The models are labeled as follows: `circ' means ``circular rotation'' (Sec.~\ref{subsec:CircularMotion}), `rad' means ``radial infall'' (Sec.~\ref{subsec:RadialMotion}), `thin' refers to a disk with opening-angle parameter $\alpha=0.1$, and `thick' to a disk with $\alpha=1$.}
\label{tbl:Parameters}
\end{table*}

\subsubsection{Infalling motion}
\label{subsec:RadialMotion}

Next, we define the radially infalling matter component.
Here, we simply assume that the motion is geodesic, with no azimuthal angular momentum and an asymptotically vanishing velocity.
As reviewed in App.~\ref{app:Flow}, the resulting four-velocity takes the form
\begin{align}
	\label{eq:RadialFourVelocity}
	u_\mathrm{rad}^\mu\pd_\mu=u_\mathrm{rad}^t\pd_t+u_\mathrm{rad}^r\pd_r+u_\mathrm{rad}^\phi\pd_\phi,
\end{align}
where
\begin{align}
	u_\mathrm{rad}^t=-g^{tt},\qquad
	u_\mathrm{rad}^r=-\sqrt{\pa{-1-g^{tt}}g^{rr}},\qquad
	u_\mathrm{rad}^\phi=-g^{t\phi}.
\end{align}
Note that the $\phi$ component is nonzero due to frame-dragging. 

\subsubsection{Combined motion}

In this paper, we only consider purely circular motion or purely radial motion, but here we show that it is easy to combine them.

Indeed, one can linearly combine these motions to obtain the total four-velocity $u^\mu\pd_\mu=u^t\pd_r+u^r\pd_r+u^\phi\pd_\phi$.
Following the notation of \citet{Pu2016}, we introduce $\Omega=u^\phi/u^t$ and write
\begin{subequations}
\begin{align}
	u^r&=(1-\beta_r)u_\mathrm{rad}^r,\\
	\Omega&=\Omega_\mathrm{circ}+(1-\beta_\phi)(\Omega_\mathrm{rad}-\Omega_\mathrm{circ}),
\end{align}
\end{subequations}
where $0\leq\beta_r\leq1$ and $0\leq\beta_\phi\leq1$ parametrize the superposition of the circular and radial components \eqref{eq:CircularFourVelocity} and \eqref{eq:RadialFourVelocity}.
Note that only $u_\mathrm{rad}^r$ is present in the first equation because our prescribed orbital four-velocity \eqref{eq:CircularFourVelocity} has vanishing radial component.
Unit-normalization fixes the time component of the four-velocity to
\begin{align}
	u^t=\sqrt{-\frac{1+g_{rr}\pa{u^r}^2}{g_{tt}+2g_{t\phi}\Omega+g_{\phi\phi}\Omega^2}},
\end{align}
and its $\phi$ component is then given by $u^\phi=\Omega u^t$.

Again, we will not examine this combined motion here, so the parameters $\beta_r$ and $\beta_\phi$ will not be discussed in the remainder of this paper---this section is simply meant to highlight that our model allows for very general flow dynamics.

\begin{figure*}
	\centering
	\includegraphics[width=0.95\textwidth]{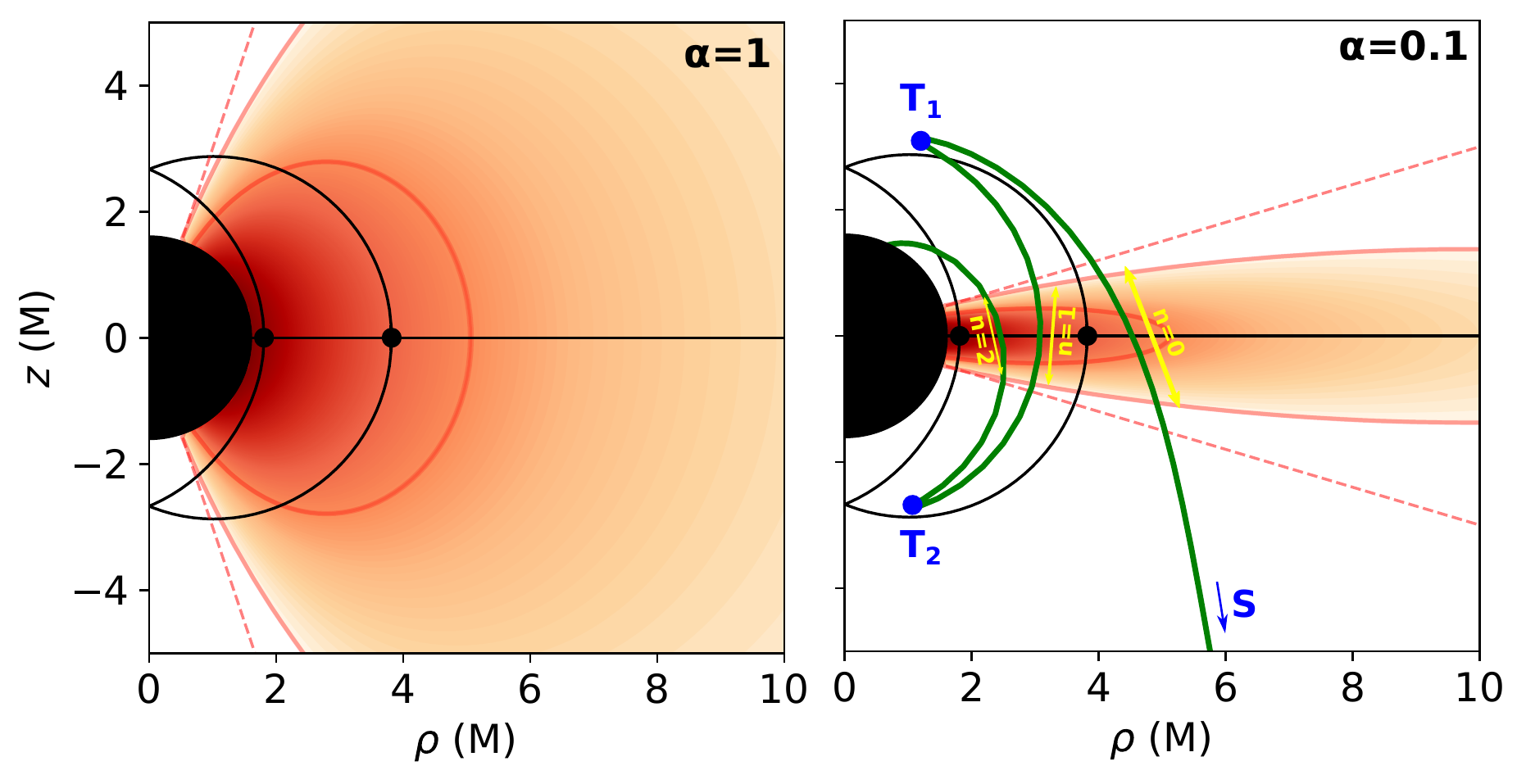} 
	\caption{Density profiles of the `thick' ($\alpha=1$, left) and `thin' ($\alpha=0.1$, right) disk models we consider. 
	The plots display the poloidal $(\rho,z)$ plane, with all azimuthal angles $\phi$ projected to one single point in the plane.
	The filled black area is the black hole event horizon.
	The solid black lines delineate the region in which bound photon orbits exist (the ``photon shell''), with the circular-equatorial orbits (prograde and retrograde) indicated by black dots.
	The red color scale encodes the log-scale profile of the electron number density, with a floor set at 1\% of the maximum density (this floor is only applied to this figure for readability; it is not applied in our model).
	The two solid red contours correspond to a density of 1\% and 10\% of the maximum.
	The dashed red lines enclose the points located at a height less than $3s_z$ above the equatorial plane, where $s_z$ is the standard deviation of the Gaussian distribution controlling the electron number density above the equator [Eq.~\eqref{eq:Prescription}].
	In the right panel, we also show a high-order null geodesic in green, with blue letters marking the distant observer's screen ($S$) and the two $\theta$ turning points ($T_1$ and $T_2$) along the geodesic.
	{Very sharp changes of direction appear at the $\theta$ turning points. These are due to the 2D projection in the $(\rho,z)$ plane of the three spatial dimensions of the null geodesic. The geodesic represented in 3D space would look perfectly smooth.}
	The $n=0$ part of the geodesic extends between $S$ and $T_1$, the $n=1$ between $T_1$ and $T_2$, and the $n=2$ between $T_2$ and the black hole.
	The yellow arrows highlight the portions of the null geodesic that are responsible for most of the $n=0$, $n=1$, and $n=2$ emission.} 
	\label{fig:DiskOrders}
\end{figure*}

\subsection{Parameter choices}

In this paper, we are primarily interested in the effects of nonzero disk thickness, the motion of the emitting material, and the black hole spin.
We thus consider extreme cases corresponding to a thin ($\alpha=0.1$) or thick disk ($\alpha=1$), purely azimuthal (Sec.~\ref{subsec:CircularMotion}) or purely radial (Sec.~\ref{subsec:RadialMotion}) motion, and low ($a=0.01$) or high ($a=0.94$) spin.
The other source parameters are fixed to take likely---or at least reasonable---values for M87*.
We choose an inclination of $i=160^\circ$ and a mass/distance pair of
$M=6.2\times 10^9\,M_\odot$ and $D=16.9$\,Mpc \citepalias{EHT5,EHT6}.
For the magnetization $\sigma$, we adopt a low value of 0.01, meaning that we consider a weakly magnetized disk.\footnote{A commonly made distinction for black hole accretion flows is that between Standard and Normal Evolution (SANE) versus a  Magnetically Arrested Disk (MAD).
Prior investigations of high-order photon rings have focused on the MAD regime \citep{Johnson2019,Chael2021}, and we are able to qualitatively reproduce these results using strongly magnetized ($\sigma=1$) thin disks ($\alpha=0.1$).
Throughout this paper, we adopt a weaker magnetization $\sigma=0.01$ corresponding to the SANE regime \citep[see, e.g., Fig.~1 of][]{Porth2019}.}
The normalizations of the electron density and temperature are chosen such that the 230\,GHz flux is of the order of 0.5\,Jy for all configurations, in accord with the analysis (and assumptions) of the 2017 EHT data \citepalias{EHT4,EHT5}.
The model is illustrated in Fig.~\ref{fig:DiskOrders}, and all the values of its parameters are given in Table~\ref{tbl:Parameters}.

\section{Image of the accretion flow} 
\label{sec:Images}

\subsection{Ray tracing and image orders}

We perform general-relativistic ray tracing in the Kerr spacetime with the model described in Sec.~\ref{sec:Model}.
We use the open-source code \href{http://gyoto.obspm.fr}{\textsc{Gyoto}} \citep{Vincent2011} to trace null geodesics backwards in time from a distant observer.
The code integrates the radiative transfer equation along the null geodesics to evolve the specific intensity $I_\nu$ across the accretion flow.
The output is an image, i.e., a map of $I_\nu$.
As discussed in App.~\ref{app:gyoto_prec}, the \textsc{Gyoto} integration parameters are chosen to ensure the highest possible numerical precision for the ray tracing and radiative transfer computations, while maintaining a reasonable computation time.

We will frequently differentiate between \textit{image orders}.
The $0^\text{th}$-order ($n=0$) primary image is defined as the image produced by selecting the part of each null geodesic that extends from the observer screen to its first angular turning point (following the ray backwards into the geometry).
The $1^\text{st}$-order ($n=1$) image is produced by selecting the part of each null geodesic that extends between its first and second angular turning points, and likewise the $2^\text{nd}$-order ($n=2$) image arises by retaining the contributions from the portion of each geodesic between its second and third angular turning points; we do not consider higher image orders.
These image orders are illustrated in Fig.~\ref{fig:DiskOrders}, and App.~\ref{app:ImageOrder} describes the technical details of their computation.

\subsection{Resulting images} 
\label{subsec:Images}

\begin{figure*}
	\centering
	\includegraphics[width=\textwidth]{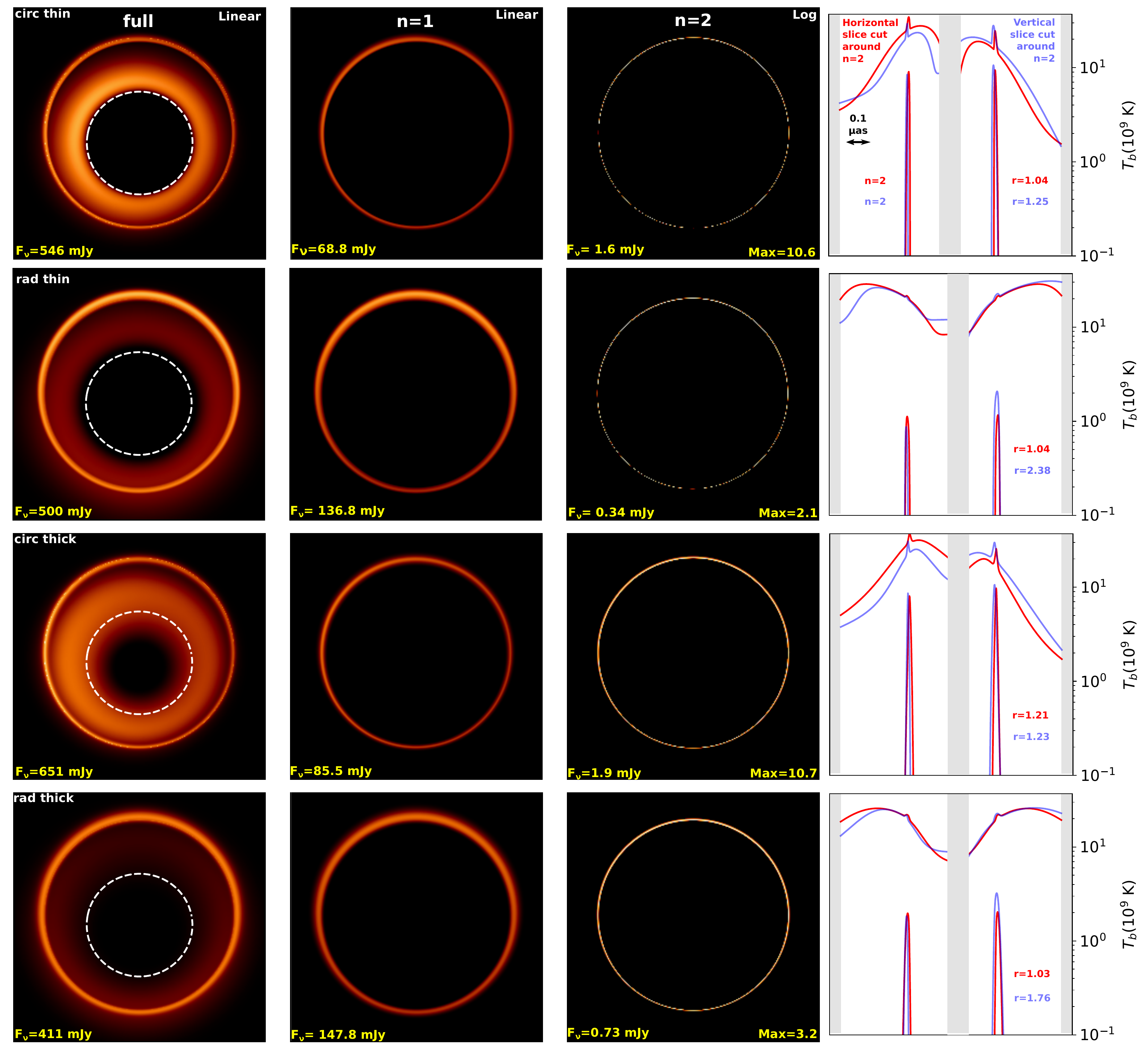} 
	\caption{Images and photon rings at 230GHz.
	\textbf{Three leftmost columns:} inner $50\,\mu$as of the \textbf{low-spin} ($a=0.01$) brightness temperature maps.
	The two left columns share the same linear color scale, which goes up to a brightness temperature of $3.7\times10^{10}$\,K.
	The third column is in logarithmic scale, with the overall scale varying across panels to enable better visualization.
	The total specific flux of each image, as well as the maximum brightness temperature of each $n=2$ image (in units of $10^9$\,K), are indicated in yellow font.
	The white dashed curve in the left column shows the primary image of the equatorial event horizon.
	\textbf{Rightmost column:} Horizontal (red) and vertical (blue) cuts of the full and $n=2$ brightness temperature profiles, centered around the $n=2$ peaks regions.
	The temperature ratios $r$ between the two $n=2$ peaks are provided.}   
	\label{fig:ImagesLowSpin}
\end{figure*}

\begin{figure*}
	\centering
	\includegraphics[width=\textwidth]{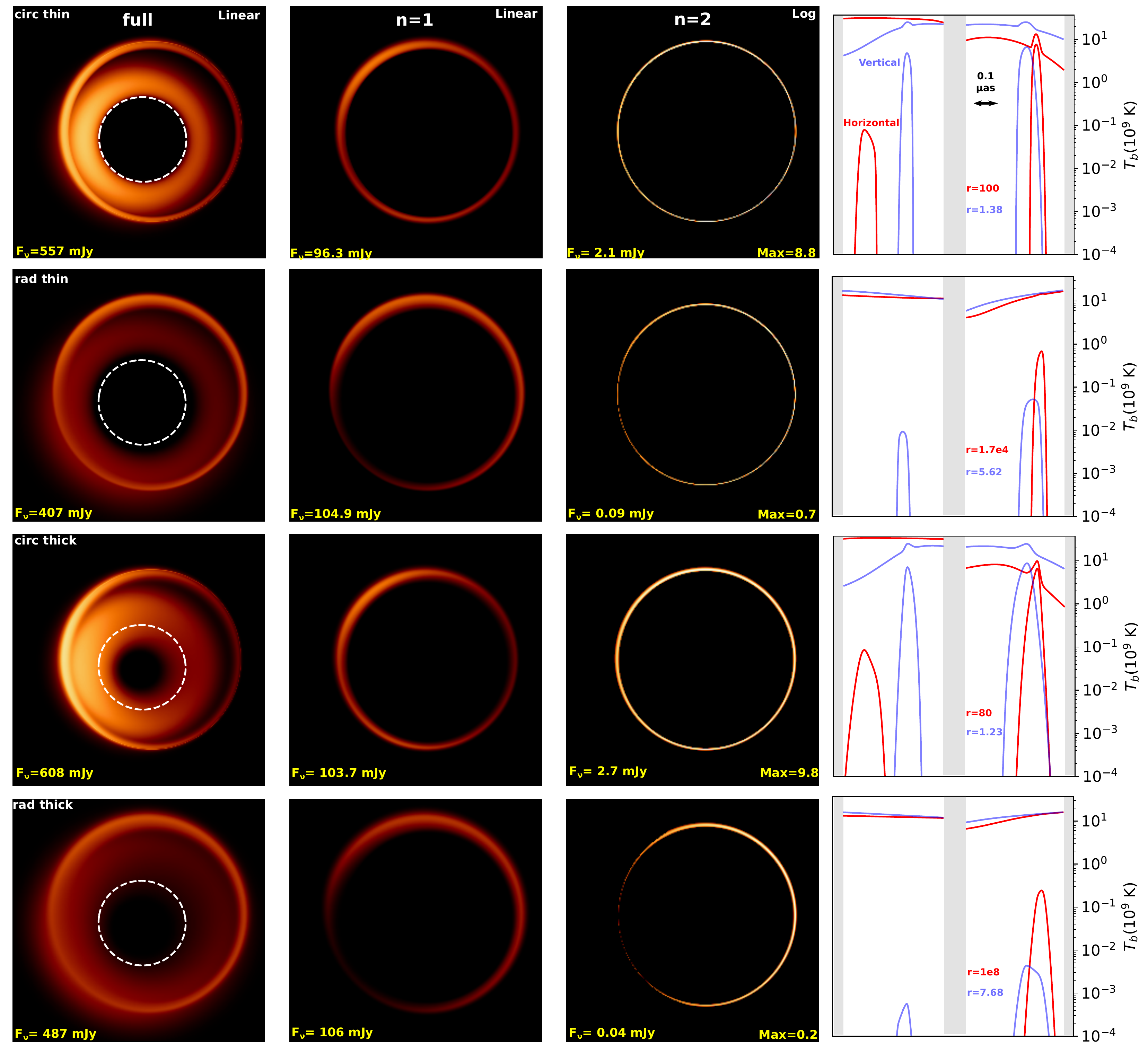} 
	\caption{Same as Fig.~\ref{fig:ImagesLowSpin} for \textbf{high spin} ($a=0.94$).
	The radial-thin and radial-thick horizontal cuts do not show the left peak of the $n=2$ ring because it is too small to be visible (the associated temperature ratios  are $r=10^4$ and $r=10^8$).
	The right column has a different scale compared to Fig.~\ref{fig:ImagesLowSpin}.}
	\label{fig:ImagesHighSpin}
\end{figure*}

The images of our models for spins $a=0.01$ and $a=0.94$ are presented in Figs.~\ref{fig:ImagesLowSpin} and \ref{fig:ImagesHighSpin}, respectively.
All of them share the same qualitative appearance, and they all display the three main features of black hole images:
\begin{itemize}
	\item[\textbullet] A central dark area, whose size depends on the astrophysical assumptions \citep[e.g.,][]{Gralla2019,Chael2021}.
	\item[\textbullet] A bright, narrow ring produced by strongly lensed photons that execute multiple orbits around the black hole on their way to the observer.
	This thin ring can be decomposed into a series of $n=1,2,\ldots$ \textit{photon rings} (often collectively referred to as ``the photon ring''), each of which is a lensed image of increasingly higher order $n$ of the accretion disk.
	These subrings, which are observable, exponentially converge to the theoretical \textit{critical curve}, which is not \citep{Bardeen1973}.
	The geometry of the critical curve is a pure function of the black hole mass and spin and of the observer inclination.
	However, the geometry and flux of observable photon rings still depend on the astrophysical assumptions \citep[e.g.,][]{Paugnat2022}.
	\item[\textbullet] A thick annular region of \textit{primary emission} (produced by $n=0$ photons travelling ``straight'' from the source to the observer, without orbiting around the black hole) that strongly depends on the astrophysical assumptions.
	In our images, it lies primarily within the thin photon ring.
\end{itemize}

For each of the models in Figs.~\ref{fig:ImagesLowSpin} and \ref{fig:ImagesHighSpin}, we present the full image (left column) and images of the $n=1$ and $n=2$ rings only (second and third columns).
The right column displays intensity along horizontal and vertical (i.e., perpendicular and parallel to the spin axis) cuts of the full image, zoomed in around the $n=2$ contribution, which is also displayed by itself.
Clearly, the details of the $n=2$ contribution are quite sensitive to the astrophysical assumptions.
We now analyze the origin of these differences.

\subsection{Intensity contribution from the \texorpdfstring{$n=2$}{n=2} ring}
\label{subsec:RingContribution}

\begin{figure*}
	\centering
	\includegraphics{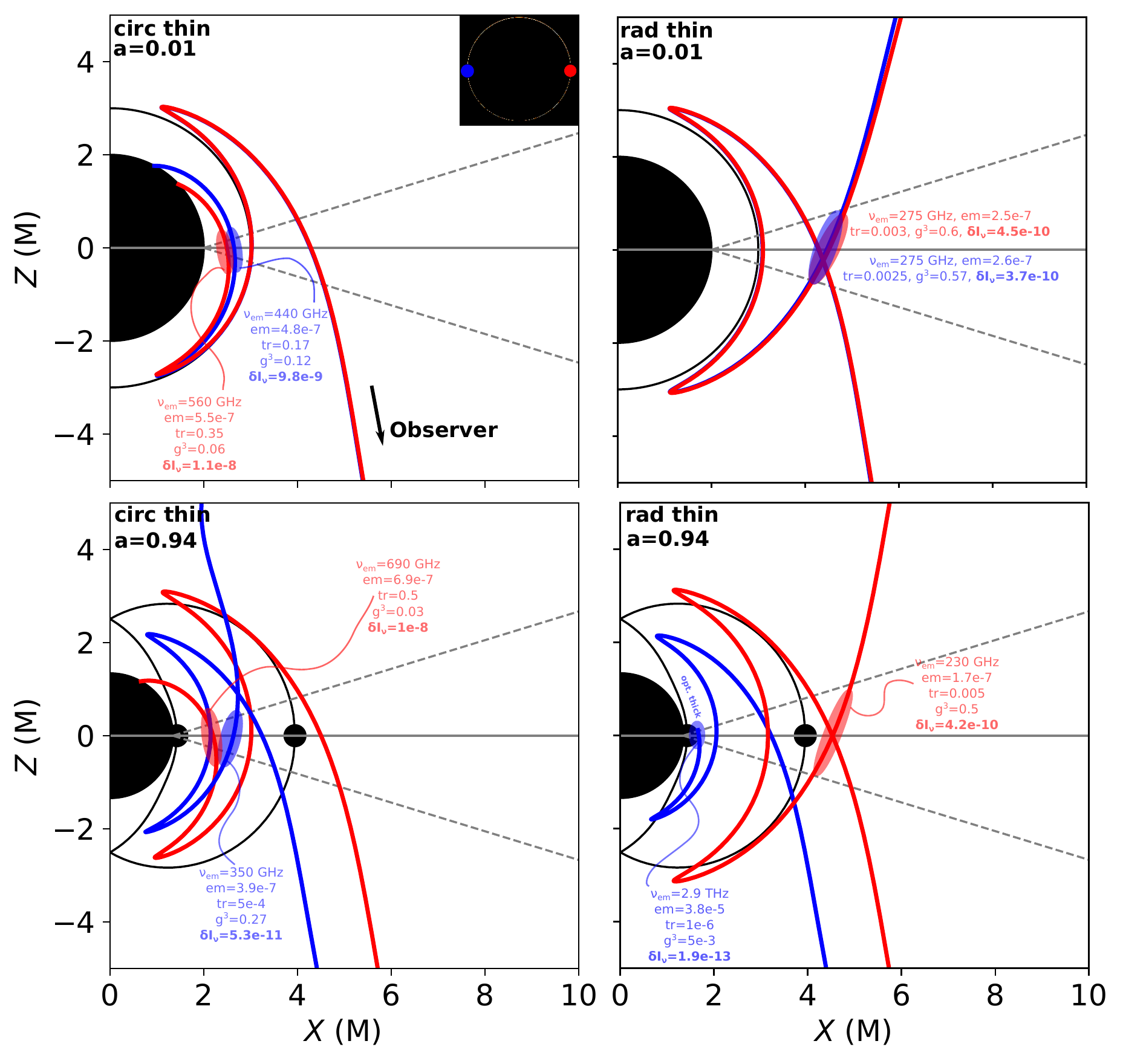}  
	\caption{Origin of $n=2$ emission.
	We show geodesics corresponding to the maximum intensity of the $n=2$ ring of each model along a horizontal cut, on the left (blue) and right (red) sides of the image (see inset of the upper-left panel).
	The accretion flow and spin parameters are specified in the upper-left corner of each panel.
	The blue and red ellipses encircle the region emitting most of the $n=2$ photons loaded onto each geodesic.
	The blue geodesic is abruptly cut in the lower-right panel because the medium becomes optically too thick (defined in the code as a transmission smaller than $10^{-6}$).
	For each ellipse, the local values of the emitted frequency ($\nu_\mathrm{em}$), self-absorbed emission (em, in cgs units), transmission (tr), and redshift factor ($g^3$) are provided, as well as the resulting increment of specific intensity $\delta I_\nu$ (see Eq.~\eqref{eq:dInu} for the definition of these quantities).} 
\label{fig:Geodesics}
\end{figure*}

Consider first the low-spin images, whose $n=2$ contributions are shown in the rightmost column of Fig.~\ref{fig:ImagesLowSpin}.
Two features are noteworthy: (i) the left and right peaks are of approximately equal intensity; (ii) the peaks in the models of radially infalling matter have lower intensity overall.
Feature (i) is a consequence of the weak frame-dragging at low spin, which implies that the left and right geodesics follow a very similar path through the flow (Fig.~\ref{fig:Geodesics} top panel).
The suppressed intensity of the radial case relative to the circular case arises from a difference in their redshift factors that can be attributed to a complicated interplay of various competing effects, which we now explore in detail.

To this end, we define a local increment of specific intensity loaded onto any given geodesic via radiative transfer.
First, we introduce a local emissivity $j_\nu$ and absorptivity $\alpha_\nu$, as well as the Planck function $B_\nu$, which depends only on the local temperature and which equals (by virtue of Kirchhoff's law) the ratio of emission to absorption.
We also need the optical depth 
\begin{align}
	\tau_\nu=\int_\text{optical path}\alpha_\nu\ed s,
\end{align}
where the integral is taken over the portion of the ray connecting the source to the observer, so that $\exp(-\tau_\nu)$ is the transmission.
The local intensity increment then reads
\begin{subequations}
\label{eq:dInu}
\begin{align}
	\dt I_\nu&=\frac{j_\nu}{\alpha_\nu}\br{1-\exp\pa{-\alpha_\nu\dt s}}\times\exp\pa{-\tau_\nu}\times g^3\\
	&=\underbrace{B_\nu\br{1-\exp\pa{-\alpha_\nu\dt s}}}_{\text{self-absorbed local emission}}\times\underbrace{\exp\pa{-\tau_\nu}}_{\text{transmission}}\times\underbrace{g^3}_{\text{redshift}}\\
	&\approx B_\nu\exp\pa{-\tau_\nu}g^3,
\end{align}
\end{subequations}
where $\dt s$ is the increment of length along the ray as measured by the emitter, and $g$ denotes the redshift factor
\begin{align}
	g=\frac{u_\mathrm{obs}\cdot p_\mathrm{obs}}{u_\mathrm{em}\cdot p_\mathrm{em}}.
\end{align}
Here, $u_\mathrm{obs/em}$ denotes the four-velocity of the observer/emitter and $p_\mathrm{obs/em}$ the photon momentum at the observer/emitter.
All the quantities appearing in Eqs.~\eqref{eq:dInu} depend on the local value of the emission frequency, which itself depends on the observed redshift, which in turn depends on the flow dynamics.

To illustrate this, we show in the top row of Fig.~\ref{fig:Geodesics} some of the geodesics that most contribute to the $n=2$ peaks visible in the rightmost column of Fig.~\ref{fig:ImagesLowSpin}.
For the circular/thin model, these light rays collect much more intensity than the analogous light rays for the radial/thin model, whose contributions are greatly suppressed by the redshift factor.
This explains point (ii) above.

To summarize, the heights of the $n=2$ peaks appearing in the rightmost columns of Figs.~\ref{fig:ImagesLowSpin} and \ref{fig:ImagesHighSpin} are controlled by three factors that contribute to the RHS of Eqs.~\eqref{eq:dInu}, namely:
\begin{enumerate}
    \item the \textit{emission}, which in our model depends mostly on radius, as it is highly concentrated near the equatorial plane (near the $n=2$ equatorial crossing of each light ray); however, it also depends on frequency and therefore redshift (see 3. below);
    \item the \textit{absorption and transmission}, which behave similarly to the emission, with the important difference that they are nonlocal, as they depend on the full history of the geodesic as it travels from the $n=2$ emission site to the observer;
    \item the \textit{redshift}, which has a complicated dependence on radius (the gravitational redshift is linked to the metric and blows up at the horizon), as well as on the relative orientation of the emitted photon and emitting particle velocity.
\end{enumerate}
The interplay of these three ingredients strongly depends on the observing frequency, on the dynamics of the flow, on the local physical conditions at the emission site (density, magnetic field, temperature), as well as on the physical conditions along the complete path followed by the photons.
It is therefore not a simple task to predict the properties of the $n=2$ contribution, and simulations are certainly the only way to investigate them.
Figure~\ref{fig:Geodesics} provides the typical values of these three quantities in the region of $n=2$ emission for each of our models.

\begin{figure*}
	\centering
	\includegraphics[width=\textwidth]{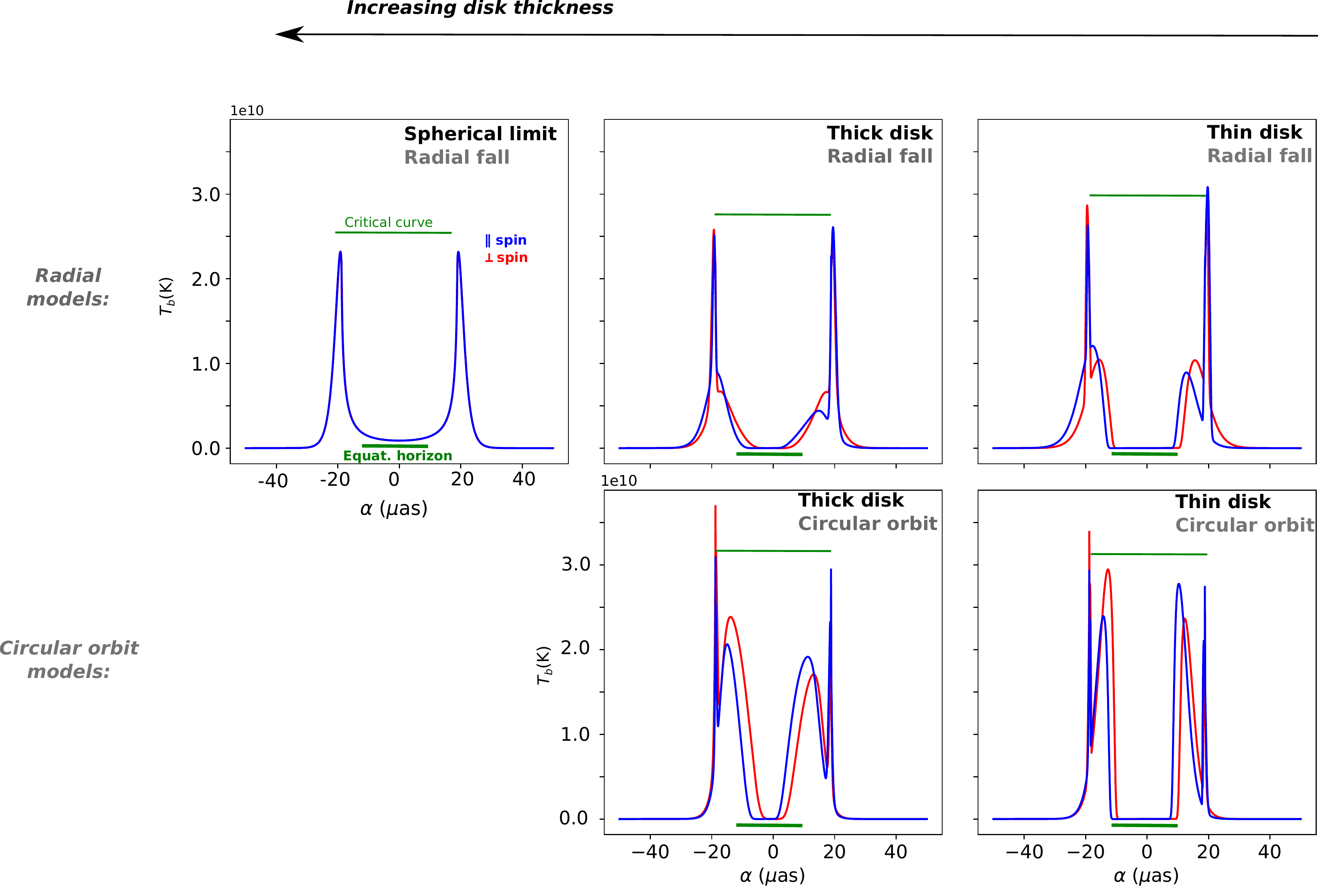}
	\caption{Intensity cuts along the directions perpendicular to spin (red) and parallel to spin (blue) for spin $a=0.01$ for the various thick-disk models used in this article (four central panels), as well as for a spherical model obtained from the thick-disk model in the limit of large disk thickness (top left panel).
	The size of the critical curve (thin line) and of the primary ($n=0$) image of the equatorial horizon (thick line) are shown in green.}
\label{fig:WeddingCakes}
\end{figure*}

At high spin, the $n=2$ peaks (shown in the rightmost column of Fig.~\ref{fig:ImagesHighSpin}) are very different from their nonrotating counterparts.
They also display two striking features: (i) in stark contrast with the nonrotating case, the left and right peaks in the horizontal cuts (perpendicular to the spin vector, shown in red) have very different heights, indicating a large brightness asymmetry in the image; (ii) this brightness asymmetry ratio (the intensity ratio of the two horizontal peaks) is extremely high for the radial models.

These features can be understood by examining the bottom row of Fig.~\ref{fig:Geodesics}, which shows that the left and right geodesics in the high-spin case probe very different parts of the inner flow, resulting in a different collected intensity.
This is closely related to the fact that as black hole spin increases, the critical curve is both distorted (from a circle) and displaced \citep[relative to the primary radiation, see, e.g.,][Fig.~5 left panel]{Chan2013}: since the $n=2$ photon ring shares these properties, the left and right geodesics visit different parts of the flow, which in turn leads to the large horizontal (perpendicular-to-spin) brightness asymmetry of the $n=2$ ring.
This explains point (i) above.

Note that for the vertical cuts (parallel to the spin vector, shown in blue in the rightmost column of Fig.~\ref{fig:ImagesHighSpin}), the brightness asymmetry is much milder.
This makes sense because at high spin, the distortion and displacement of the critical curve and $n=2$ ring primarily act in the direction orthogonal to the spin.

The extreme horizontal brightness asymmetry obtained in models of radial infall can be attributed to a strong absorption of the $n=2$ radiation along one of the geodesics, as well as to an extreme redshift effect due to the fact that the $n=2$ emission is loaded onto the left geodesic (in blue in Fig.~\ref{fig:Geodesics} lower-right panel) very close to the event horizon.
This explains point (ii) above.

\subsection{Intensity cuts}
\label{subsec:IntensityCuts}

We conclude this section by revisiting the framework proposed in the introduction, in which realistic models are regarded as lying on a continuum between the shadow and wedding cake extremes.
To this end, here we discuss intensity cuts of all the models analyzed in this paper, starting with the extreme models (spherical and equatorial) and then moving on to our thick-disk models.

\subsubsection{Extreme models: spherical \& equatorial}

Figure~\ref{fig:Menu} of the introduction presents the intensity cuts of two analytical models that are not based on the thick-disk model that we develop here.
These two models are defined as follows:
\begin{itemize}
	\item[\textbullet] The \textit{spherical infall} model consists of a spherically symmetric distribution of emitting matter located everywhere outside the event horizon.
	This matter is falling into the black hole with four-velocity defined in Sec.~\ref{subsec:RadialMotion}.
	The number density and temperature of the emitting fluid are chosen at the event horizon and evolve with radius following the power laws defined in Eqs.~\eqref{eq:Prescription} (i.e., $n_\mathrm{e}(r)\propto r^{-2}$, $T_\mathrm{e}(r)\propto r^{-1}$).
	The magnetic field is prescribed by demanding that the magnetization $\sigma$ defined in Eq.~\eqref{eq:Magnetization} remain constant, at the same value as in our thick-disk model, $\sigma=0.01$.
	We consider thermal synchrotron emission and ignore self-absorption.
    \item[\textbullet] The \textit{equatorial orbiting} model is that
of \citetalias{Gralla2020} with an ad-hoc emission profile (their emission model 1).
	There is also no absorption considered in this model.
\end{itemize}

\subsubsection{Canonical thick-disk models}

We now discuss the intensity cuts of our four canonical thick-disk models, which are displayed in Fig.~\ref{fig:WeddingCakes}.
We focus on the case of low spin, as the high-spin case is qualitatively similar.

The ``spherical limit'' in the top-left panel is the $\alpha\to\infty$ limit of our disk model with radially infalling matter.
This large-thickness limit of our thick-disk model exactly coincides with the spherical infall model discussed in the previous subsection.
The only difference is that in the upper-left panel of Fig.~\ref{fig:WeddingCakes}, we take synchrotron self-absorption into account.
This spherical limit is similar to the models used by \citet{Falcke2000} and \citet{Narayan2019}, except that we self-consistently include absorption in our model of synchrotron emission.  

The $\alpha\to0$ limit of our thick-disk is numerically tricky, and the emission profile of the analytical equatorial model depicted in Fig.~\ref{fig:Menu} differs from that obtained with our thick-disk models.
For this reason, we do not show an ``equatorial limit'' intensity cut in Fig.~\ref{fig:WeddingCakes}.

Figure~\ref{fig:WeddingCakes} shows a continuum of models spanning a large parameter space, from near-equatorial to spherical emission, and from circularly orbiting matter to radially infalling matter.
We highlight a few important observations:
\begin{itemize}
    \item[\textbullet] The radial spherical model (top left) shows a prominent photon ring surrounding a large shadow that essentially occupies the full interior of the critical curve.
    The existence of the shadow is due to the strong redshift effect on the radiation emitted by matter that falls towards the black hole.\footnote{Null geodesics arriving inside the critical curve have no radial turning points so the emitted radiation is directed towards increasing radii, while the emitter is heading towards decreasing radii, giving rise to strong redshift suppression of the observed intensity via the ``headlight'' effect.}
    The spherical model is the only case where the intensity profile does not decompose into discrete layers ($n=0,1,2$).
    \item[\textbullet] Our thick-disk models produce a wide range of profiles that interpolate between the spherical limit and the equatorial model depicted in Fig.~\ref{fig:Menu}.
    Like the equatorial model, these models all produce intensity profiles that decompose into distinct $n=0$ and $n=1$ contributions.
    The $n=2$ photon ring is also clearly present for all the circular models; in the radial models, it is also present but (as discussed above) weaker and less easily discerned.
    The contrast between the photon rings and the $n=0$ emission is stronger for the radial models because their $n=0$ contribution is strongly suppressed (by the same redshift effect that produces a shadow in the spherical model).
    As a consequence, the central brightness depression is larger in the radial models than in the circular models; for the latter, as for the equatorial model, the central dark region is restricted to the interior of the apparent equatorial horizon.
\end{itemize}

\section{Visibility signatures}
\label{sec:Visibilities}

One of the motivations for this paper is the prospect of an experimental detection of the $n=2$ photon ring via its long-baseline ``ringing'' in the Fourier plane, which might be observable with future space-based VLBI missions \citepalias[\citealp{Johnson2019,Pesce2019,Gralla2020b,GrallaLupsasca2020};][]{Gralla2020}.
To check whether such a feature would be observable in our models, we compute the visibility along cuts in the Fourier plane in the directions parallel and perpendicular to the spin axis.  
Following \citetalias{Gralla2020}, we leverage the projection-slice theorem to simplify the calculation.
For a given orientation, we determine the line integrals on all lines perpendicular to the orientation (a 1D cut of the Radon transform of the image), and we Fourier transform this 1D function to find the desired cut of the 2D Fourier transform, which is the radio visibility.
The magnitude of this 1D complex visibility is the visibility amplitude, which we will simply refer to as the visibility.  
In the same way that we extracted from the full image its $n=1$ and $n=2$ components, we will compute the full Fourier transform and its $n=1$ and $n=2$ components.
Our Fourier transform conventions agree with \citetalias{Gralla2020}.

Table~\ref{tbl:FFT} shows the main features of the visibilities associated to the various models considered in this study.
In particular, it provides the visibility amplitude level at 40 and 100\,G$\lambda$, as well as the baseline thresholds (and corresponding visibility levels) past which the $n=1$ and $n=2$ rings start to dominate the signal.
These thresholds were called $b_1$ and $b_2$ in \citet{Paugnat2022}.

The following criterion is used to define when the $n^\text{th}$ ring starts to dominate the visibility.
The full image decomposes into layers indexed by $n$, $I_\nu(\mathbf{x})=I_0(\mathbf{x})+I_1(\mathbf{x})+I_2(\mathbf{x})+\ldots$, and the full visibility $V(\mathbf{u})=V_0(\mathbf{u})+V_1(\mathbf{u})+V_2(\mathbf{u})+\ldots$ does too, by linearity of the Fourier transform.
We consider a sliding baseline window $u\sim u_w$ of fixed width 10\,G$\lambda$, which is approximately twice the photon ring diameter.
At each angle $\varphi$ in the baseline plane, we compute over this window $u\sim u_w$ the mean visibility $\av{\ab{V(u,\varphi)}}_w$ of the full image $I_\nu(\mathbf{x})$ and the mean visibility $\av{\ab{V_n(u,\varphi)}}_w$ of the $n^\text{th}$ ring image $I_n(\mathbf{x})$ only.
We define the baseline threshold $b_n(\varphi)$ past which the $n^\text{th}$ ring dominates the visibility to be the shortest baseline window $u_w\equiv b_n(\varphi)$ such that the percentage difference between these two means dips below a given threshold $p$:
\begin{align}
	\label{eq:Percent}
	\frac{\av{\ab{V_n(u,\varphi)}}_w-\av{\ab{V(u,\varphi)}}_w}{\av{\ab{V(u,\varphi)}}_w}<p
	\Longrightarrow
	\begin{array}{c}
		\text{$n^\text{th}$ ring dominates}\\
		\text{on }u\gtrsim b_n(\varphi)\equiv u_w.
	\end{array}
\end{align}
We will generally take $p=5\%$, unless otherwise stated.

\begin{table*}[h!]
\centering
\begin{tabular}{|c|c|c|c|c|}
	\hline
	Model & 40\,G$\lambda$  & 100\,G$\lambda$ & $n=1$ & $n=2$ \\
	\hline\hline
	$a=0.01$ circular thin &
	\diagbox{\makecell{10\,mJy\\$\,$}}{\makecell{$\,$\\9.2\,mJy}} &
	\diagbox{\makecell{2.7\,mJy\\$\,$}}{\makecell{$\,$\\2.4\,mJy}} &
	\diagbox{\makecell{(\textcolor{red}{140\,G$\lambda$},\\\textcolor{red}{2\,mJy})}}{\makecell{(150\,G$\lambda$,\\1\,mJy)}} &
	\diagbox{\makecell{(1300\,G$\lambda$,\\20\,$\mu$Jy)}}{\makecell{(1400\,G$\lambda$,\\25\,$\mu$Jy)}} \\
	\hline
	$a=0.01$ radial thin &
	\diagbox{\makecell{12\,mJy\\$\,$}}{\makecell{$\,$\\14.4\,mJy}} &
	\diagbox{\makecell{3.3\,mJy\\$\,$}}{\makecell{$\,$\\2.2\,mJy}} &
	\diagbox{\makecell{(65\,G$\lambda$,\\7.5\,mJy)}}{\makecell{(54\,G$\lambda$,\\10\,mJy)}} &
	\diagbox{\makecell{(900\,G$\lambda$,\\5.6\,$\mu$Jy)}}{\makecell{(800\,G$\lambda$,\\10\,$\mu$Jy)}} \\
	\hline
	$a=0.01$ circular thick &
	\diagbox{\makecell{7.7\,mJy\\$\,$}}{\makecell{$\,$\\8.6\,mJy}} &
	\diagbox{\makecell{2.8\,mJy\\$\,$}}{\makecell{$\,$\\2.4\,mJy}} &
	\diagbox{\makecell{(44\,G$\lambda$,\\7.7\,mJy)}}{\makecell{(54\,G$\lambda$,\\6.8\,mJy)}} &
	\diagbox{\makecell{(650\,G$\lambda$,\\43.2\,$\mu$Jy)}}{\makecell{(650\,G$\lambda$,\\55\,$\mu$Jy)}} \\
	\hline
	$a=0.01$ radial thick &
	\diagbox{\makecell{11.2\,mJy\\$\,$}}{\makecell{$\,$\\12.2\,mJy}} &
	\diagbox{\makecell{2.1\,mJy\\$\,$}}{\makecell{$\,$\\2.2\,mJy}} &
	\diagbox{\makecell{(34\,G$\lambda$,\\12.8\,mJy)}}{\makecell{(35\,G$\lambda$,\\14.4\,mJy)}} &
	\diagbox{\makecell{(660\,G$\lambda$,\\14.9\,$\mu$Jy)}}{\makecell{(660\,G$\lambda$,\\23\,$\mu$Jy)}} \\
	\hline\hline
	$a=0.94$ circular thin &
	\diagbox{\makecell{11.2\,mJy\\$\,$}}{\makecell{$\,$\\12.0\,mJy}} &
	\diagbox{\makecell{1.8\,mJy\\$\,$}}{\makecell{$\,$\\2.0\,mJy}} &
	\diagbox{\makecell{(N/A,\\N/A)}}{\makecell{(\textcolor{red}{112\,G$\lambda$},\\\textcolor{red}{1.8\,mJy})}} &
	\diagbox{\makecell{(N/A,\\N/A)}}{\makecell{(700\,G$\lambda$,\\57\,$\mu$Jy)}} \\
	\hline
	$a=0.94$ radial thin &
	\diagbox{\makecell{7.1\,mJy\\$\,$}}{\makecell{$\,$\\10.4\,mJy}} &
	\diagbox{\makecell{1.4\,mJy\\$\,$}}{\makecell{$\,$\\1.3\,mJy}} &
	\diagbox{\makecell{(54\,G$\lambda$,\\5.5\,mJy)}}{\makecell{(44.2\,G$\lambda$,\\9.7\,mJy)}} &
	\diagbox{\makecell{(N/A,\\N/A)} }{\makecell{(896\,G$\lambda$,\\0.25\,$\mu$Jy)}} \\
	\hline
	$a=0.94$ circular thick &
	\diagbox{\makecell{7.3\,mJy\\$\,$}}{\makecell{$\,$\\10.7\,mJy}} &
	\diagbox{\makecell{0.7\,mJy\\$\,$}}{\makecell{$\,$\\1.8\,mJy}} &
	\diagbox{\makecell{(63.8\,G$\lambda$,\\3.5\,mJy)}}{\makecell{(\textcolor{red}{54\,G$\lambda$},\\\textcolor{red}{6\,mJy})}} &
	\diagbox{\makecell{(N/A,\\N/A)}}{\makecell{(328\,G$\lambda$,\\180\,$\mu$Jy)}} \\
	\hline
	$a=0.94$ radial thick &
	\diagbox{\makecell{6.2\,mJy\\$\,$}}{\makecell{$\,$\\7.4\,mJy}} &
	\diagbox{\makecell{0.5\,mJy\\$\,$}}{\makecell{$\,$\\0.6\,mJy}} &
	\diagbox{\makecell{(34.4\,G$\lambda$,\\7.3\,mJy)}}{\makecell{(34.4\,G$\lambda$,\\10.0\,mJy)}} &
	\diagbox{\makecell{(N/A,\\N/A)}}{\makecell{(847\,G$\lambda$,\\0.01\,$\mu$Jy)}} \\
	\hline 
\end{tabular}  
\caption{Key features of the visibility amplitudes of the various models considered at $\boldsymbol{\nu_\mathrm{obs}=230}$\,\textbf{GHz}: for each model, we list the visibility amplitude $\ab{V(u,\varphi)}$ at $u=40$\,G$\lambda$ and 100\,G$\lambda$, as well as the baseline thresholds $b_1(\varphi)$ and $b_2(\varphi)$ (and the corresponding visibility amplitudes) past which the $n=1$ and $n=2$ contributions start to dominate the signal [Eq.~\eqref{eq:Percent}].
Each box contains two sets of numbers separated by a diagonal: these correspond to a baseline polar angle $\varphi=0^\circ$ (perpendicular to spin, lower left) or $\varphi=90^\circ$ (aligned with spin, upper right).
The percent level $p$ of Eq.~\eqref{eq:Percent} is 5\% everywhere, except for the numbers in red, where it is set to 10\% (in these cases, the ring never dominates according to the 5\% criterion).
The `N/A' in the $n=2$ column for the high-spin cases at $\varphi=0^\circ$ mean that the $n=2$ signal never dominates even for $p=10\%$.}
\label{tbl:FFT}
\end{table*}

\begin{figure*}
	\centering
	\includegraphics[height=0.95\textheight]{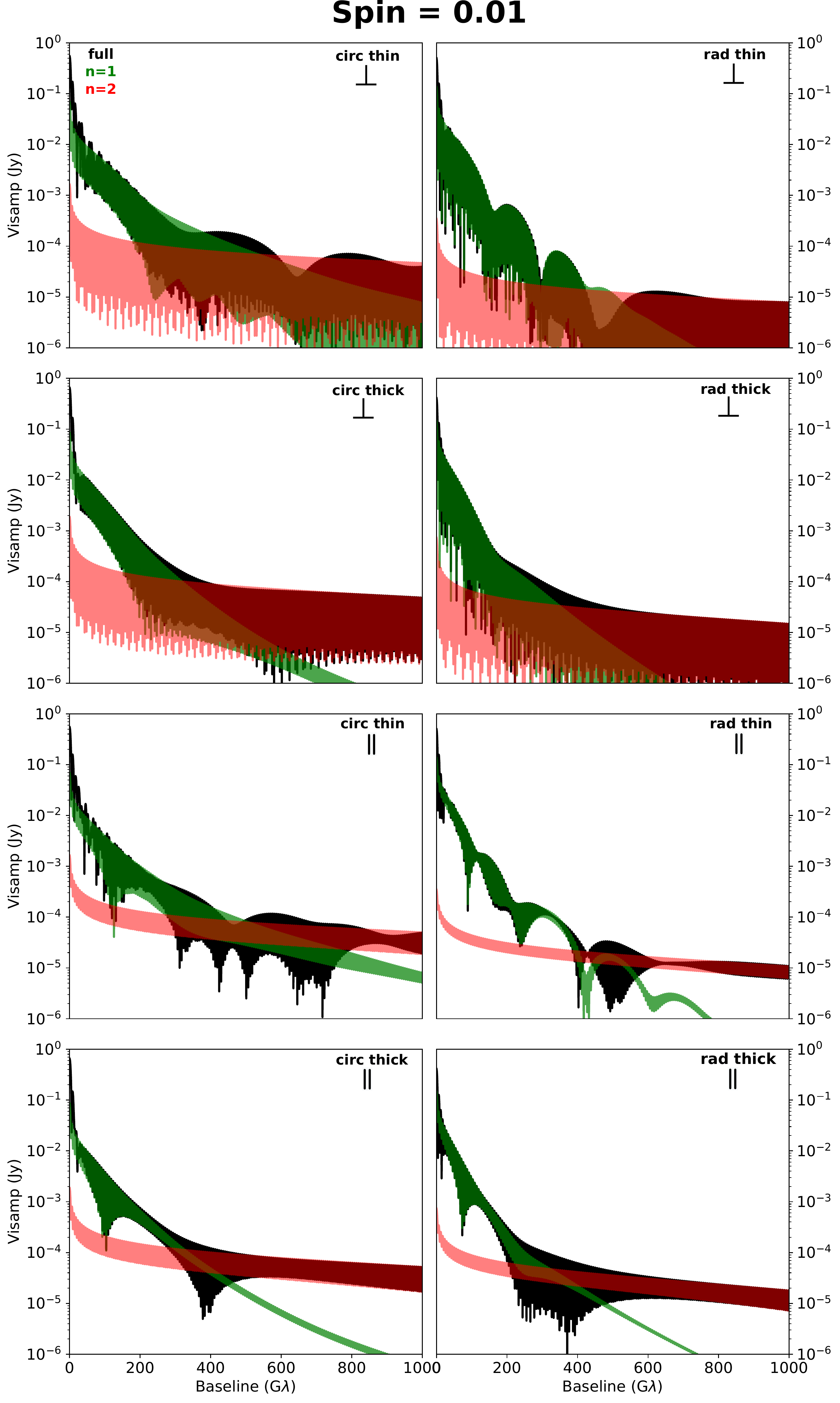} 
	\caption{Visibility amplitude profiles for all \textbf{low-spin} ($a=0.01$) models.
	The full profile $\ab{V(u,\varphi)}$ is shown in black, the $n=1$ profile $\ab{V_1(u,\varphi)}$ in green, and the $n=2$ profile $\ab{V_2(u,\varphi)}$ in red.
	The four upper panels correspond to an orientation $\varphi=0^\circ$ (perpendicular to the spin axis, $\perp$), and the four lower panels to an orientation $\varphi=90^\circ$ (parallel to the spin axis, $\parallel$).
	The accretion flow model is specified in the top-right corner of each panel.}
	\label{fig:VisampsLowSpin} 
\end{figure*}

We again focus our comments on the $n=2$ ring signature.
In all low-spin models, the $n=2$ ring clearly dominates the signal on sufficiently long baselines $u\gtrsim b_2$, with the threshold $b_2$ ranging from $\sim600-1400$\,G$\lambda$.
The corresponding visibility amplitude displays clear oscillations at levels varying between $\sim10-50\,\mu$Jy (see Table~\ref{tbl:FFT} and Fig.~\ref{fig:VisampsLowSpin}).
Thus, these low-spin models match the idealized models considered in \citetalias{Gralla2020} and \citet{Paugnat2022} reasonably well. 

\begin{figure*} 
	\centering
	\includegraphics[height=0.97\textheight]{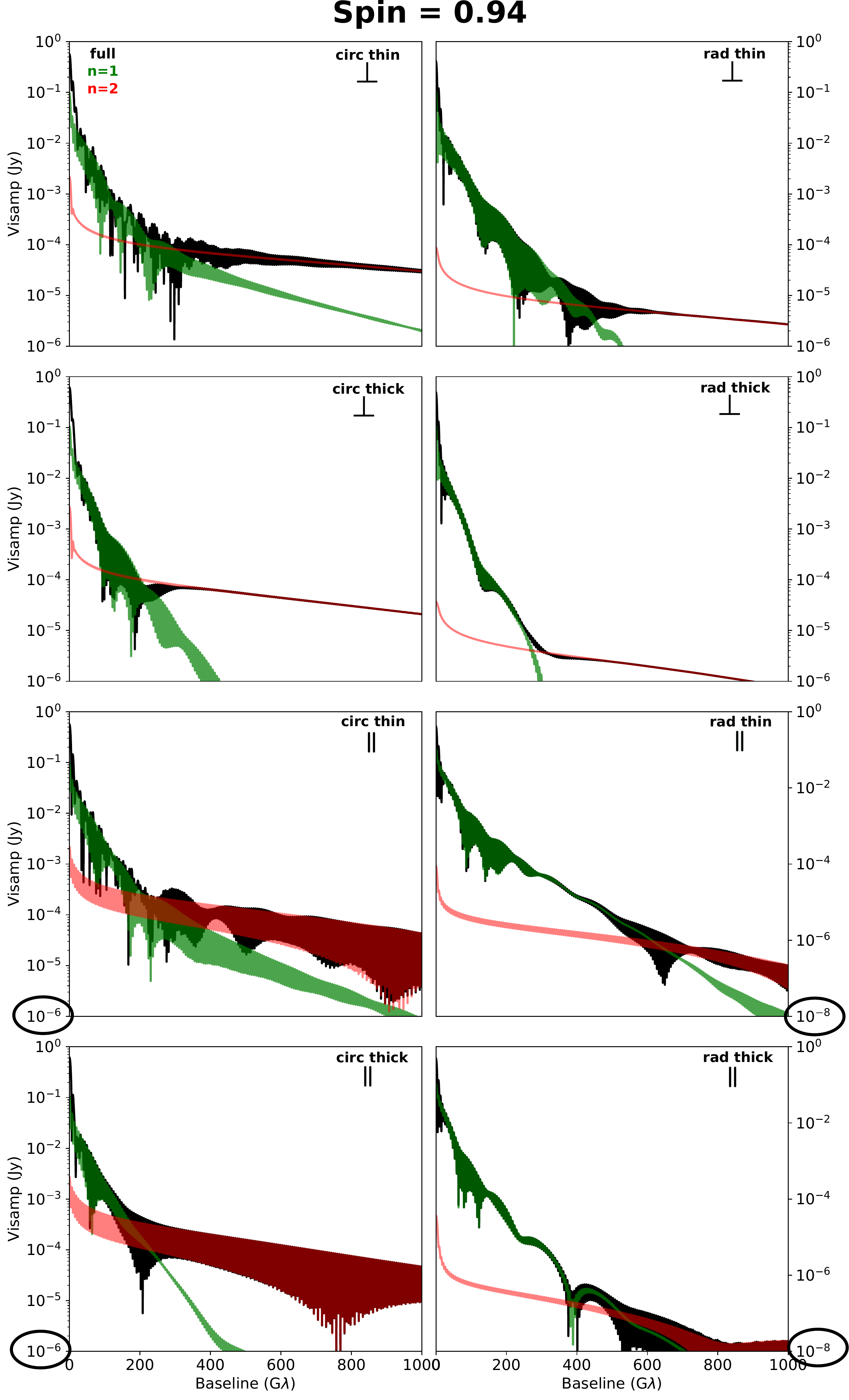} 
	\caption{Same as Fig.~\ref{fig:VisampsLowSpin} for the \textbf{high-spin} ($a=0.94$) models.
	In all models, the $n=2$ signal is very weak at $\varphi=0^\circ$ (orientation perpendicular to the spin axis) due to the strong frame-dragging effect at high spin (see text for details).
	Note the different vertical scale for the two lower right panels.}
	\label{fig:VisampsHighSpin}
\end{figure*}

At high spin, by contrast, it is only on the baseline oriented parallel to the spin axis ($\varphi=90^\circ$) that the $n=2$ ring clearly dominates the signal (see the four lower panels of Fig.~\ref{fig:VisampsHighSpin}).
However, even at this orientation, the models of radially infalling matter produce a very weak visibility amplitude in the $n=2$-dominated regime $\approx0.01-0.1\,\mu$Jy.
This is directly related to the low height of the $n=2$ peaks in the rightmost column of Fig.~\ref{fig:ImagesHighSpin}: the blue profiles for the two models of radial infall display a contrast with the $n=1$ radiation of $\approx 100-1000$.

On baselines perpendicular to the spin axis ($\varphi=0^\circ$), all the high-spin models produce a very weak $n=2$ signal, with essentially no oscillation at long baselines (see four upper panels of Fig.~\ref{fig:VisampsHighSpin}).
This is the reason why some numbers are missing in Table~\ref{tbl:FFT}: these entries correspond to cases in which the $n=2$ oscillation is vanishingly small.
To understand this effect, recall that the oscillations in visibility amplitude are caused by ``interference'' between opposite peaks in the image, with the amplitude of the oscillation set by the smaller of the two \citep[Eq.~(7) in][]{Gralla2020b}.
Given that all high-spin, spin-perpendicular cuts display a large brightness asymmetry between the two $n=2$ peaks (see the red profiles of the rightmost column in Fig.~\ref{fig:ImagesHighSpin}), the resulting visibility cuts exhibit essentially no oscillations associated with the $n=2$ ring.
This asymmetry between the $n=2$ peaks is itself due to the interplay between absorption and redshift effects on the various parts of the $n=2$ ring (see Sec.~\ref{subsec:RingContribution}), as illustrated by the geodesics plotted in the bottom row of Fig.~\ref{fig:Geodesics}.

To summarize what we have discussed so far, the $n=2$ peaks in the intensity cuts displayed in the rightmost column of Figs~\ref{fig:ImagesLowSpin} and \ref{fig:ImagesHighSpin} have two features of fundamental importance: (i) the ratio of flux in the $n=2$ image $I_2(\mathbf{x})$ relative to the total flux in the full image $I_\nu(\mathbf{x})$ controls the average strength of the visibility in the $n=2$-dominated regime; (ii) the brightness asymmetry (intensity ratio) between the left and right (or top and bottom) peaks in the $n=2$ image cuts controls the amplitude of the visibility oscillations in the $n=2$-dominated regime.
These empirical observations are in perfect agreement with the analytical study of~\citet{Gralla2020b}.
They allow one to guess whether a model will produce a strong $n=2$ signal by simply computing a high-resolution intensity cut along a given orientation in the image plane, which is very cheap in terms of computational resources.

\begin{table*}[h!]
\centering
\begin{tabular}{|c|c|c|c|c|}
	\hline
	Model & 40\,G$\lambda$ & 100\,G$\lambda$ & $n=1$ & $n=2$ \\
	\hline\hline
	$a=0.94$ circular thin &
	\diagbox{\makecell{30.1\,mJy\\$\,$}}{\makecell{$\,$\\11.6\,mJy}} &
	\diagbox{\makecell{6.4\,mJy\\$\,$}}{\makecell{$\,$\\2.5\,mJy}} &
	\diagbox{\makecell{(\textcolor{red}{83\,G$\lambda$},\\\textcolor{red}{8.6\,mJy})}}{\makecell{(\textcolor{red}{112\,G$\lambda$},\\\textcolor{red}{1.8\,mJy})}} &
	\diagbox{\makecell{(593\,G$\lambda$,\\150\,$\mu$Jy)}}{\makecell{(455\,G$\lambda$,\\176\,$\mu$Jy)}} \\
	\hline
	$a=0.94$ radial thin &
	\diagbox{\makecell{24.2\,mJy\\$\,$}}{\makecell{$\,$\\15.7\,mJy}} &
	\diagbox{\makecell{3.6\,mJy\\$\,$}}{\makecell{$\,$\\4.1\,mJy}} &
	\diagbox{\makecell{(34.4\,G$\lambda$,\\26.3\,mJy)}}{\makecell{(34.4\,G$\lambda$,\\18.5\,mJy)}} &
	\diagbox{\makecell{(397\,G$\lambda$,\\110\,$\mu$Jy)}}{\makecell{(377\,G$\lambda$,\\209\,$\mu$Jy)}} \\
	\hline
	$a=0.94$ circular thick &
	\diagbox{\makecell{18.6\,mJy\\$\,$}}{\makecell{$\,$\\7\,mJy}} &
	\diagbox{\makecell{4.9\,mJy\\$\,$}}{\makecell{$\,$\\1.5\,mJy}} &
	\diagbox{\makecell{(\textcolor{red}{45\,G$\lambda$},\\\textcolor{red}{18\,\,mJy})}}{\makecell{(\textcolor{red}{45\,G$\lambda$},\\\textcolor{red}{5.4\,mJy})}} &
	\diagbox{\makecell{(254\,G$\lambda$,\\372\,$\mu$Jy)}}{\makecell{(384\,G$\lambda$,\\140\,$\mu$Jy)}} \\
	\hline
	$a=0.94$ radial thick &
	\diagbox{\makecell{19.4\,mJy\\$\,$}}{\makecell{$\,$\\15.3\,mJy}} &
	\diagbox{\makecell{1.6\,mJy\\$\,$}}{\makecell{$\,$\\3.5\,mJy}} &
	\diagbox{\makecell{(35\,G$\lambda$,\\23.2\,mJy)}}{\makecell{(35\,G$\lambda$,\\17.4\,mJy)}} &
	\diagbox{\makecell{(374\,G$\lambda$,\\100\,$\mu$Jy)}}{\makecell{(284\,G$\lambda$,\\227\,$\mu$Jy)}} \\
	\hline
\end{tabular}  
\caption{Same as Table~\ref{tbl:FFT}, but at $\boldsymbol{\nu_\mathrm{obs}=345}$\,\textbf{GHz}.
Only high-spin models are considered.}
\label{tbl:FFT345}
\end{table*}

\begin{figure}
	\centering
	\includegraphics[width=\columnwidth]{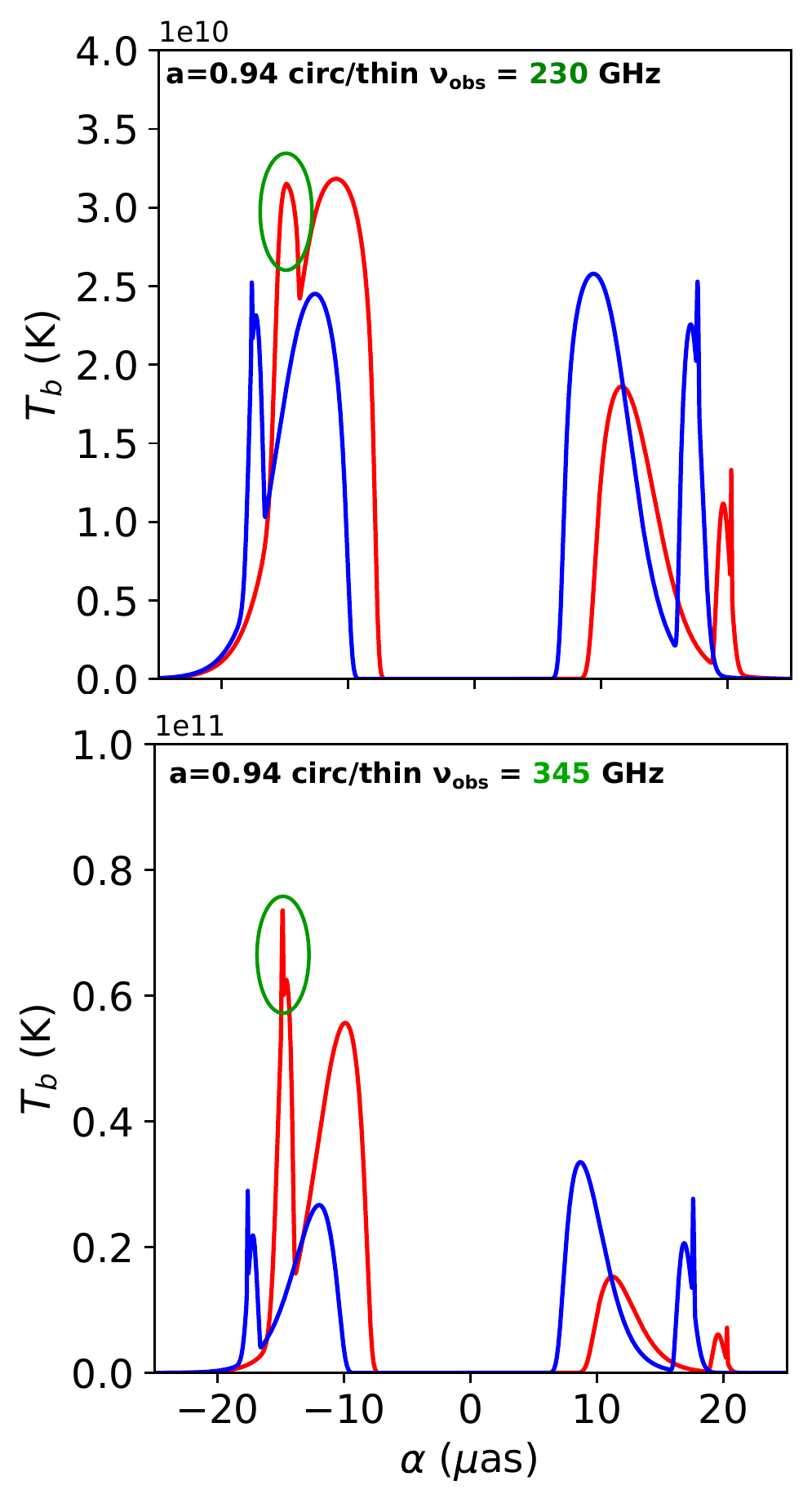}
	\caption{Intensity cuts along directions perpendicular (red) and parallel (blue) to spin for the circular/thin model at high spin, for two observing frequencies: 230\,GHz and 345\,GHz.
 	The emission region is more optically thin at 345\,GHz, so the higher-order peaks are more pronounced.
 	In particular, the $n=2$ left peak of the spin-perpendicular case is very sharp at
 	$345$~GHz, and fully absorbed at $230$~GHz
 	(see green ellipses).}
	\label{fig:FrequencyComparison}
\end{figure}

The fact that the $n=2$ signal is absent from the spin-perpendicular visibility in the high-spin models is problematic from an experimental perspective.
However, since absorption is the main culprit in suppressing the $n=2$ contribution, one might expect the signal to reappear at higher observation frequencies, where absorption is lower.
While we have so far restricted our attention to the present-EHT observation frequency $\nu_\mathrm{obs}=230$\,GHz, we can also consider a higher frequency $\nu_\mathrm{obs}=345$\,GHz of planned future VLBI observations \citepalias[\citealt{ngEHT}; ][]{EHT2}.
Table~\ref{tbl:FFT345} provides the same information as Table~\ref{tbl:FFT}, for the high-spin models observed at $\nu_\mathrm{obs}=345$\,GHz.
This higher frequency contains a clear $n=2$ signal in all cases, and the associated visibility level is in the regime $\sim100\,\mu$Jy that is considered potentially observable by \citetalias{Gralla2020}.  
The stronger $n=2$ signal at higher frequency is also readily seen in the image cross-sections.
For example, Fig.~\ref{fig:FrequencyComparison} displays intensity cuts for the circular/thin model at these two frequencies, and the higher-order peaks are clearly much sharper at 345\,GHz.

\section{Conclusion}
\label{sec:Conclusion}

In this article, we studied the images and visibility amplitude profiles of a variety of models of geometrically thin and thick disks, including the effects of absorption.
We found that the size of the central brightness depression depends strongly on astrophysical assumptions, and that the photon ring is generically discrete rather than smooth, supporting the ``wedding cake'' heuristic over the ``black hole shadow'' paradigm.
We also studied the long-baseline visibility amplitude, focusing on the signature of the $n=2$ photon ring of M87*.
Although the $n=2$ signal is always clear at low spin at 230\,GHz, it can disappear at high spin due to a combination of frame-dragging and absorption along the line of sight.
However, at 345\,GHz, we find a clear $n=2$ signal even at high spin, because the absorption is less prominent at this higher frequency.  

We thus find that 345\,GHz is a promising target for future space-VLBI observations of the photon ring of M87*.
However, we emphasize that this conclusion applies only for the particular astrophysical conditions that we have considered herein.
Alternative reasonable choices of inclination, total compact flux, magnetization, density profile, etc., may very well lead to a different conclusion.
We also cannot discount the possibility that the source geometry is qualitatively different, consisting for instance of a tilted disk \citep[e.g.,][]{White2020}, a jet-dominated profile \citep[e.g.,][]{Kawashima2021}, or emission from current sheets \citep[e.g.,][]{Crinquand2022}.
Finally, we should keep in mind that the source itself may be variable in ways that we do not understand and cannot predict.
For example, the total compact source flux density appears to have varied by a factor of 2 across a decade of VLBI observations, which would influence the density scale and hence the system optical depth \citep[]{Wielgus2020}. 
It is therefore of paramount importance to extend our analysis to a larger astrophysical parameter space covering all reasonable possibilities for the source.

\section*{Acknowledgements}

{F.H.V thanks F. Roy from Paris Observatory/LUTH for his help in compiling the \textsc{Gyoto} code on the MesoPSL cluster.
We thank Alejandro C\'ardenas-Avenda\~no, Hadrien Paugnat, and George Wong for helpful comments and discussions.
George Wong also provided the data for the green curve in Fig.~\ref{fig:Menu}, and Alejandro C\'ardenas-Avenda\~no computed the equatorial horizon images in Figs.~\ref{fig:ImagesLowSpin} and \ref{fig:ImagesHighSpin} using \texttt{AART} \citep{CardenasAvendano2022}.

\noindent This work was supported in part by NSF grant PHY–1752809 to the University of Arizona.
It was also granted access to the HPC resources of MesoPSL financed by the R\'egion \^Ile-de-France and the project Equip@Meso (reference ANR-10-EQPX-29-01) of the Programme d'Investissements d'Avenir supervised by the Agence Nationale pour la Recherche.
Lastly, A.L. also gratefully acknowledges support from Will and Kacie Snellings, while M.W. thanks Alexandra Elbakyan for her contributions to the open science initiative.}

\bibliography{M87}
\bibliographystyle{aa}

\appendix

\section{Relativistic thermal synchrotron emission and absorption}
\label{app:Synchrotron}

The synchrotron power per unit frequency emitted by a single ultrarelativistic electron is well-known \citep[see][Sec.~6.2]{Rybicki1979} to be
\begin{align}
	\label{eq:SynchrotronPower}
	\frac{\ed E}{\ed t\ed\nu}=\sqrt{3}\frac{e^3B\sin{\theta}}{mc^2}F\pa{\frac{\nu}{\nu_\mathrm{crit}}},
\end{align}
where
\begin{align}
	\nu_\mathrm{crit}=\frac{3}{2}\gamma^2\nu_\mathrm{cyclo}\sin{\theta},\qquad
	\nu_\mathrm{cyclo}=\frac{eB}{2\pi mc}.
\end{align}
Here, $\gamma$ denotes the Lorentz factor of the electron, $\theta$ the angle between the direction of emission and the magnetic field, and
\begin{align}
	F(x)\equiv x\int_x^\infty K_{5/3}(u)\ed u
	=
	\begin{cases}
	   2^{2/3}\Gamma\pa{\frac{2}{3}}x^{1/3}&\text{for }x\ll1,\\
	   \sqrt{\frac{\pi x}{2}}e^{-x}&\text{for }x\gg1,
	\end{cases}
\end{align}
with $K_{5/3}(u)$ a modified Bessel function of the second kind.
Note that in these formulas, we have identified the pitch angle between the magnetic field and the direction of the electron's velocity with the emission angle $\theta$, which is a valid approximation for ultrarelativistic electrons \citep{Rybicki1979}.
Also note that the derivation of Eq.~\eqref{eq:SynchrotronPower} is essentially built on the strong beaming effect of the gyrating ultrarelativistic electrons, which is thus taken into account from the very start in this treatment.

The associated thermal synchrotron emissivity is
\begin{align}
	\label{eq:SynchrotronEmissivityDefinition}
	j_\nu&\equiv\frac{\ed E}{\ed t\ed\nu\ed V\ed\Omega}
	=\frac{\sqrt{3}}{4\pi}\frac{e^3B\sin{\theta}}{mc^2}\int_0^\infty F\pa{\frac{\nu}{\nu_\mathrm{crit}}}\frac{\ed n_\mathrm{e}}{\ed\gamma}\ed\gamma,
\end{align}
where
\begin{align}
	\frac{\ed n_\mathrm{e}}{\ed\gamma}=\frac{n_\mathrm{e}}{\Theta_\mathrm{e}}\frac{\gamma\pa{\gamma^2-1}^{1/2}}{K_2\pa{\Theta_\mathrm{e}^{-1}}}\exp\pa{-\frac{\gamma}{\Theta_\mathrm{e}}}
\end{align}
is the relativistic thermal (Maxwell-J\"uttner) distribution, $K_2(u)$ is a modified Bessel function of the second kind, $\Theta_\mathrm{e}=k_BT_\mathrm{e}/mc^2$ is the dimensionless electron temperature, and the factor of $\pa{4\pi}^{-1}$ implicitly assumes an isotropic distribution of the electrons' momenta.
Note that the integral over $\gamma$ in Eq.~\eqref{eq:SynchrotronEmissivityDefinition} is evaluated over the range $[0,\infty)$, even though physically, $\gamma\geq1$.
Extending the range of integration to include the interval $[0,1]$ simplifies the computation of the integral, while maintaining a good accuracy (as we will see below).
Using the asymptotic expansions of $F(x)$ for $x\ll1$ and $x\gg1$, the emissivity \eqref{eq:SynchrotronEmissivityDefinition} may be analytically expressed as \citep[see, e.g.,][]{Leung2011}
\begin{align}
	j_\nu\approx
    \begin{cases}
       \displaystyle\frac{2^{4/3}\pi}{3}\frac{n_\mathrm{e}e^2\nu_s}{c\Theta_\mathrm{e}^2}X^{1/3}&\text{for }X\ll1,\\
     \displaystyle n_\mathrm{e}\frac{\sqrt{2}\pi e^2\nu_s}{6c\Theta_\mathrm{e}^2}Xe^{-X^{1/3}}&\text{for }X\gg 1,
    \end{cases}
\end{align}
where
\begin{align}
	\label{eq:DefinitionX}
	X=\frac{\nu}{\nu_s},\qquad
	\nu_s=\frac{2}{9}\nu_\mathrm{cyclo}\Theta_\mathrm{e}^2\sin{\theta}.
\end{align}
\citet{Leung2011} also provide a fitting function that bridges these two asymptotic regimes:
\begin{align}
	\label{eq:ApproximateSynchrotronEmissivity}
	j_\nu^\mathrm{approx}=n_\mathrm{e}\frac{\sqrt{2}\pi e^2\nu_s}{3cK_2\pa{\Theta_\mathrm{e}^{-1}}}\pa{X^{1/2}+2^{11/12}X^{1/6}}^2e^{-X^{1/3}}.
\end{align}
To summarize, Eq.~\eqref{eq:ApproximateSynchrotronEmissivity} is derived under the following list of approximations: we assume the electrons are ultrarelativistic and their population isotropic, we slightly extend the integration bounds on $\gamma$, and we match the asymptotic expansions of $F(x)$.

In our ray tracing code, we use the formula \eqref{eq:ApproximateSynchrotronEmissivity} for the emissivity, averaged over the emission direction $\theta$:
\begin{align}
	\av{j_\nu^\mathrm{approx}}=\frac{1}{4\pi}\int j_\nu^\mathrm{approx}\ed\Omega
	=\frac{1}{2}\int_0^\pi j_\nu^\mathrm{approx}\sin{\theta}\ed\theta,
\end{align}
where $\ed\Omega=\sin{\theta}\ed\theta\ed\phi$ and the emissivity is independent of the azimuthal angle $\phi$.

The exact expression for the emissivity, free from the above approximations, is (see \citealp{Leung2011} for details)
\begin{align}
	\label{eq:ExactSynchrotronEmissivity}
	j_\nu^\mathrm{exact}=\frac{2\pi e^2\nu}{c\ab{\cos{\theta}}}\sum_{n=n_-}^\infty\int_{\gamma_-(n)}^{\gamma_+(n)}\frac{1}{\beta}\mathcal{J}_n(\gamma)\frac{\ed n_\mathrm{e}}{\ed\gamma}\ed\gamma,
\end{align}
where $n_-=\frac{\nu}{\nu_\mathrm{cyclo}}\ab{\sin{\theta}}$ and
\begin{align}
	\gamma_\pm(n)=\frac{n\frac{\nu_\mathrm{cyclo}}{\nu}\pm\ab{\cos{\theta}}\sqrt{\pa{n\frac{\nu_\mathrm{cyclo}}{\nu}}^2-\sin^2{\theta}}}{\sin^2{\theta}},
\end{align}
where $n$ is the integer that is being summed over in Eq.~\eqref{eq:ExactSynchrotronEmissivity}.
The dimensionless function $\mathcal{J}_n(\gamma)$ is given in terms of the Bessel function of the first kind $J_n(z)$ and its derivative $J_n'(z)$ by
\begin{subequations}
\begin{gather}
	\mathcal{J}_n(\gamma)=\br{MJ_n(z)}^2+\br{NJ_n'(z)}^2,\\
	M=\frac{\cos{\theta}-\beta\cos{\alpha_0}}{\sin{\theta}},\qquad
	z=\frac{\nu\gamma\beta}{\nu_\mathrm{cyclo}}\sin{\theta}\sin{\alpha_0},\\
	N=\beta\sin{\alpha},\qquad
	\cos{\alpha_0}=\frac{1}{\beta\cos{\theta}}\pa{1-\frac{n}{\gamma}\frac{\nu_\mathrm{cyclo}}{\nu}}.
\end{gather}
\end{subequations}
The sum over $n$ in Eq.~\eqref{eq:ExactSynchrotronEmissivity} accounts for the helical motion of electrons along magnetic field lines, which makes each electron emit at every multiple of its Doppler-shifted gyrofrequency.

The accuracy of the approximation \eqref{eq:ApproximateSynchrotronEmissivity} can be assessed by comparing it with the exact expression \eqref{eq:ExactSynchrotronEmissivity}.
This is done in Figs.~10 and 11 of \citet{Leung2011}, where the relative error between these two quantities at $\theta=30^\circ$ and $80^\circ$ is displayed as a function of $\Theta_\mathrm{e}$ and $\nu/\nu_\mathrm{cyclo}$.
This error is typically of the order of a few percent for $\Theta_\mathrm{e}\in[1,10]$ and $\nu/\nu_\mathrm{cyclo}\in[10^3,10^5]$, which includes the typical ranges for these parameters in the inner region of the accretion flow surrounding M87*.
Indeed, the temperature of the source is $T_\mathrm{e}\approx10^{11}$\,K, so $\Theta_\mathrm{e}\approx15$ (and the accuracy gets better with increasing $\Theta_\mathrm{e}$), while the magnetic field is $B\approx 10$\,G, which implies a cyclotron frequency $\nu_\mathrm{cyclo}\approx10^7$\,Hz, so that $\nu/\nu_\mathrm{cyclo}\approx10^4$.
On the other hand, the error typically exceeds 10\% when $\Theta_\mathrm{e}\lesssim1$, which means that the approximation \eqref{eq:ApproximateSynchrotronEmissivity} does not accurately model the emission from the outer region of the accretion flow.
Nevertheless, this is not an issue for us because these contributions are negligible compared to the emission from the inner region of the flow.

The thermal absorption coefficient is simply related to the emissivity by Kirchhoff's law $\alpha_\nu=j_\nu/B_\nu$, where $B_\nu$ is the Planck function, so the approximations for the absorption are exactly the same as for the emission.

\section{Synchrotron emission profile for M87*}
\label{app:Emissivity}

\begin{figure}
	\centering
	\includegraphics[width=\columnwidth]{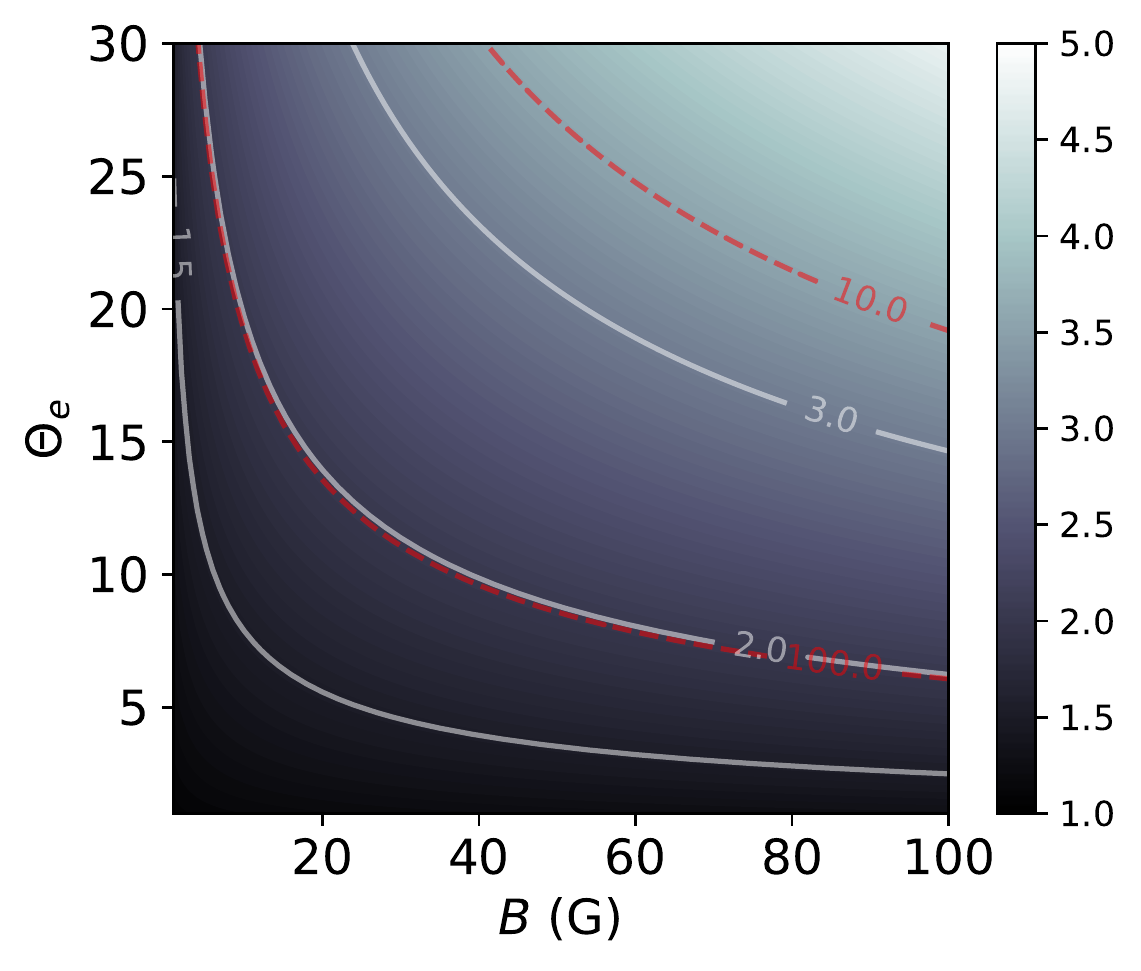} 
	\caption{Ratio of $(X^{1/2}+2^{11/12}X^{1/6})^2$ to $X$ [or equivalently, of the emissivity \eqref{eq:HighTemperatureEmissivity} to its approximation \eqref{eq:ApproximateEmissivity}] as a function of magnetic field strength $B$ and dimensionless temperature $\Theta_\mathrm{e}$, fixing $\sin{\theta}=1$ and $\nu=230$\,GHz (and neglecting redshift effects).
	The solid white curves are level sets of ratios of 1.5, 2 and 3.
	The dashed red curves are level sets of the $X$ parameter corresponding to $X=10$ and $X=100$.} 
	\label{fig:X}
\end{figure}

\subsection{Thermal synchrotron}

Since $\Theta_\mathrm{e}\approx15$ for M87*, we may assume that $\Theta_\mathrm{e}\gg1$ for this source.
In that case, the approximation \eqref{eq:ApproximateSynchrotronEmissivity} for the emissivity can be further simplified using $K_2(x)\stackrel{x\to0}\approx2x^{-2}$.
This substitution results in the further approximation
\begin{align}
	\label{eq:HighTemperatureEmissivity}
	j_\nu\stackrel{\Theta_\mathrm{e}\gg1}{\approx}n_\mathrm{e}\frac{\sqrt{2}\pi e^2\nu_\mathrm{cyclo}\sin{\theta}}{27c}\pa{X^{1/2}+2^{11/12}X^{1/6}}^2e^{-X^{1/3}},
\end{align}
which is accurate to better than 10\% provided that $\Theta_\mathrm{e}>1.5$, and to better than 1\% provided $\Theta_\mathrm{e}>5$.

Plugging typical values for M87* yields an estimate for $X$ of
\begin{align}
	\label{eq:X}
	X\approx\frac{1600}{\sin{\theta}}\pa{\frac{\nu[\mathrm{Hz}]}{10^{12}}}\pa{\frac{10}{B [\mathrm{G}]}}\pa{\frac{10}{\Theta_\mathrm{e}}}^2,
\end{align}
where the frequency $\nu$ and magnetic field strength $B$ should be expressed in Hertz and Gauss, respectively.
The factor of $1600$ incorporates all the constant terms in the definition \eqref{eq:DefinitionX} of $X$.

Since this expression satisfies $X\gg1$ for all typical values of M87* parameters, we can further simplify the approximation \eqref{eq:HighTemperatureEmissivity} for the emissivity by taking its $X\to\infty$ limit.
In the regime $X\gg1$, we may replace $(X^{1/2}+2^{11/12}X^{1/6})^2$ by $X$ to find
\begin{align}
	\label{eq:ApproximateEmissivity}
	j_\nu\stackrel{\Theta_\mathrm{e},X\gg1}{\approx}n_\mathrm{e}\frac{\sqrt{2}\pi e^2\nu_\mathrm{cyclo}\sin{\theta}}{27c}Xe^{-X^{1/3}},
\end{align}
which is a slight underestimation because $(X^{1/2}+2^{11/12}X^{1/6})^2$ is always bigger than $X$.
Their precise ratio is plotted as a function of $B$ and $\Theta_\mathrm{e}$ in Fig.~\ref{fig:X} and is always of order $\sim1-10$ for typical parameter values for M87*.

Next, since we are only interested in the overall profile of $j_\nu$, we may drop all constant prefactors and write
\begin{align}
	\label{eq:EmissivityProfile}
	j_\nu\propto\frac{n_\mathrm{e}\nu}{\Theta_\mathrm{e}^2}e^{-X^{1/3}}
	=\frac{n_\mathrm{e}\nu}{\Theta_\mathrm{e}^2}\exp\br{-\mathcal{C}\pa{\frac{\nu}{B\Theta_\mathrm{e}^2}}^{1/3}},
\end{align}
where $\mathcal{C}=(9\pi mc/e)^{1/3}$ is a constant.

Since we are assuming that the density and temperature fall off as $r^{-2}$ and $r^{-1}$, respectively, we may immediately conclude that the radial dependence of the emissivity is simply given by
\begin{align}
	j_\nu(r)\propto\exp\br{-X(r)^{1/3}},
\end{align}
provided that we neglect the radial dependence of the frequency, which is affected by a position-dependent redshift; this is only a reasonable assumption provided the emission is not too close to the horizon $r=r_+$, where the redshift diverges.

Returning to the estimate \eqref{eq:X} for $X$, and using the assumed magnetic field fall-off $B\propto r^{-1}$ [Eq.~\eqref{eq:Magnetization}], we may now write
\begin{align}
	X\approx\frac{3.7\times10^5}{B_\mathrm{inner}\Theta_\mathrm{e;inner}^2\sin{\theta}}\pa{\frac{r}{r_\mathrm{inner}}}^3,
\end{align}
where we have set $\nu=230$\,GHz and once again neglected the frequency redshift.
Here, $r_\mathrm{inner}$ is the innermost radius of the disk, while $B_\mathrm{inner}$ and $\Theta_{e,\mathrm{inner}}$ respectively denote the innermost values of the magnetic field and dimensionless temperature.
Hence,
\begin{align}
	\label{eq:Zeta}
	j_\nu(r)\propto\exp\br{-\zeta\frac{r}{r_\mathrm{inner}}},\qquad
	\zeta=\pa{\frac{3.7\times10^5}{B_\mathrm{inner}\Theta_\mathrm{e;inner}^2\sin{\theta}}}^{1/3}.
\end{align}

\begin{figure}
	\centering
	\includegraphics[width=\columnwidth]{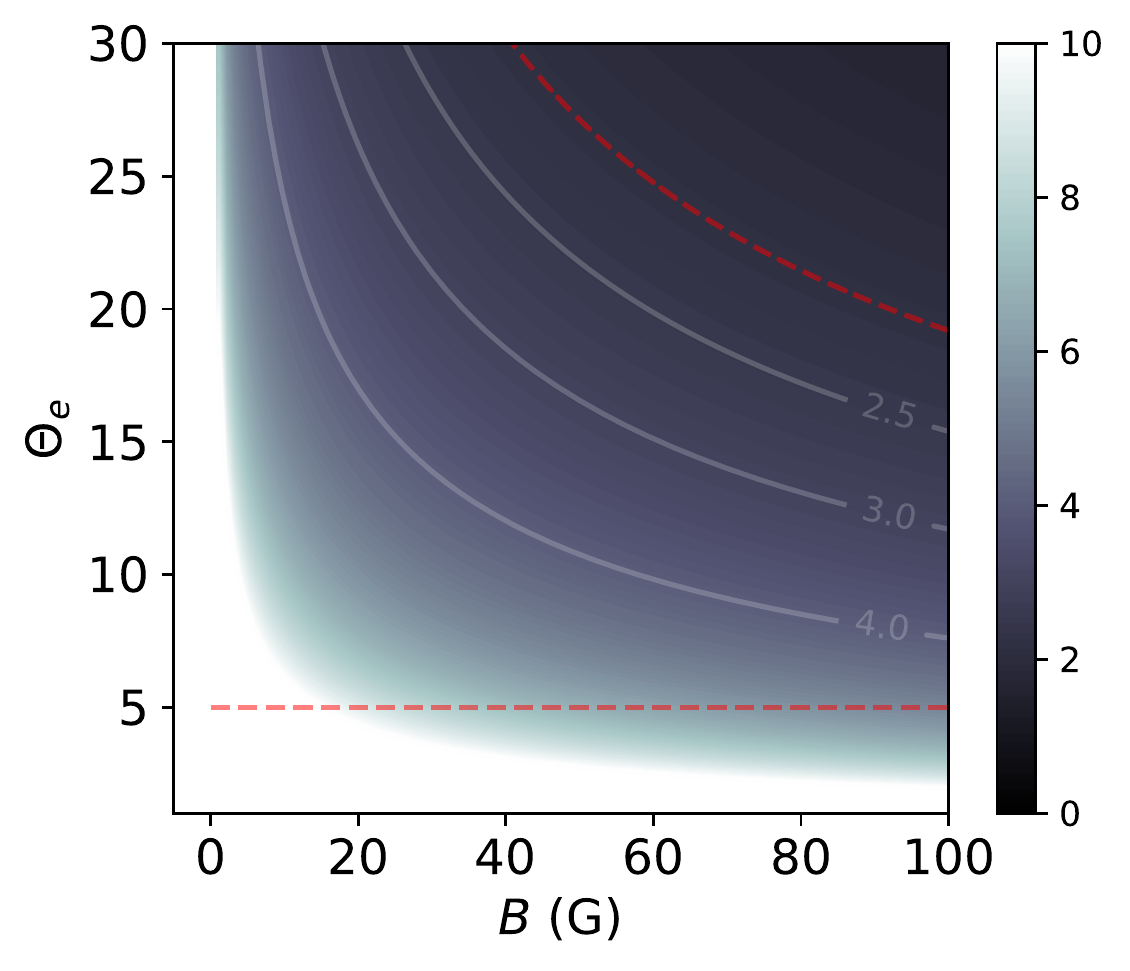} 
	\caption{Values of the parameter $\zeta$ [Eq.~\eqref{eq:Zeta}] as a function of the magnetic field strength $B$ and dimensionless temperature $\Theta_\mathrm{e}$, fixing $\sin{\theta}=1$.
	Our models always stay in the part of the plane located above the dashed horizontal red line ($\Theta_\mathrm{e}>5$, where the approximation \eqref{eq:HighTemperatureEmissivity} is good) and below the dashed red contour ($X>10$, where the approximation \eqref{eq:ApproximateEmissivity} is excellent).
	The solid white contours are level sets of $\zeta$.} 
\label{fig:Zeta}
\end{figure}

To summarize, Eq.~\eqref{eq:Zeta} is derived under the approximations listed below Eq.~\eqref{eq:ApproximateSynchrotronEmissivity}, together with the following additional assumptions: we assume that $\Theta_\mathrm{e}\gg1$ and $X\gg1$ (which puts us between the red contours of Fig.~\ref{fig:Zeta}), and we prescribe power-law fall-offs for the density, temperature, and magnetic field as described in our model ($n_\mathrm{e}\propto r^{-2}$, $\Theta_\mathrm{e}\propto r^{-1}$, $B\propto r^{-1}$).
Should the various power-law indices differ from this choice, one could derive a corresponding correction to Eq.~\eqref{eq:Zeta} starting from the more general expression \eqref{eq:EmissivityProfile}.
We reiterate that this approximation underestimates the emissivity by a factor of a few, and that its quality increases with $X$ (Fig.~\ref{fig:X}).

The thermal-synchrotron emission profile for typical M87* parameters thus decays exponentially in the radius, with a slope $\zeta$ depending on the exact conditions.
Table~\ref{tbl:Zeta} lists the values of $\zeta$ for our geometrically thin models, and Fig.~\ref{fig:Zeta} displays $\zeta$ as a function of the magnetic field strength and temperature.

For thermal synchrotron emission, the absorptivity may be obtained from Kirchhoff's law as
\begin{align}
	\alpha_\nu=\frac{j_\nu}{B_\nu(T_\mathrm{e})},
\end{align}
where $B_\nu(T_\mathrm{e})$ is the Planck function at the local electron temperature $T_\mathrm{e}$.
At 230\,GHz, photons are deep in the Rayleigh-Jeans regime where $h\nu\ll k_BT_\mathrm{e}$, so that this relation simplifies to
\begin{align}
	\alpha_\nu\approx\frac{c^2}{2k_BT_\mathrm{e}\nu^2}j_\nu.
\end{align}
Assuming as usual a temperature profile $T\propto r^{-1}$, we obtain
\begin{align}
	\label{eq:Absorptivity}
	\alpha_\nu(r)\propto\frac{r}{r_\mathrm{inner}}\exp\br{-\zeta\frac{r}{r_\mathrm{inner}}}.
\end{align}
The highly lensed emission originates from small radii $r\gtrsim r_\mathrm{inner}$, for which the exponential profiles \eqref{eq:Zeta} and \eqref{eq:Absorptivity} provide good approximations to the emissivity and absorptivity, respectively.

\begin{table}[h!]
	\centering
	\begin{tabular}{|c|c|} 
		\hline
		Model & $\zeta$ \\
		\hline
		$a=0.01$ circ/thin & 4.0 \\
		$a=0.01$ rad/thin & 3.4 \\
		$a=0.94$ circ/thin & 3.4 \\
		$a=0.94$ rad/thin & 2.7 \\
		\hline
	\end{tabular}
	\caption{Values of the parameter $\zeta$ in a few of our models, for which the emissivity profile is approximately $\exp(-\zeta r/r_\mathrm{inner})$, provided that $r$ is far enough from the event horizon and redshift effects are neglected.}
	\label{tbl:Zeta}
\end{table}

\subsection{Power-law synchrotron}

Although we only study thermal-synchrotron emission in this paper, it is interesting to also consider a nonthermal population of electrons (accelerated by magnetic reconnection, for instance) with synchrotron emission following a power law with index $p$.
Still following \citet{Leung2011}, the emissivity is then
\begin{align}
	j_\nu^\mathrm{PL}&=n_\mathrm{e}\frac{e^2\nu_\mathrm{cyclo}}{c}\frac{3^{p/2}(p-1)\sin{\theta}}{2(p+1)\pa{\gamma_\mathrm{min}^{1-p}-\gamma_\mathrm{max}^{1-p}}}\notag\\
	&\quad\times\Gamma\pa{\frac{3p-1}{12}}\Gamma\pa{\frac{3p+19}{12}}\pa{\frac{\nu}{\nu_\mathrm{cyclo}}}^{-(p-1)/2},
\end{align}
where $\gamma_\mathrm{max/min}$ are the maximum/minimum values of the Lorentz factor for electrons in the distribution and $\Gamma(x)$ the gamma function.
Following the same steps as in the previous section leads to
\begin{align}
	j_\nu^\mathrm{PL}\propto\pa{\frac{r}{r_\mathrm{inner}}}^{-(p+5)/2},
\end{align}
i.e., a power-law emissivity profile depending only on the power-law index of the electron distribution, and not on the physical conditions of the flow.
For the particular value $p=3$, $j_\nu^\mathrm{PL}\propto r^{-2}$.

The power-law absorptivity $\alpha_\nu^\mathrm{PL}$ is also provided by \citet{Leung2011} and also follows a power law that depends only on the power-law index of the electron distribution.
For $p=3$, it behaves as $\alpha_\nu^\mathrm{PL}\propto r^{-9/2}$.
Thus, it appears that the radial profile of the emissivity and absorptivity can vary significantly with the choice of electron distribution, and this could in turn have significant impact on the $n=2$ ring signature (see Sec.~\ref{subsec:Images}).
It would therefore be useful to also investigate this signature in models with nonthermal emission.

\section{Geometrically thick accretion flows}
\label{app:Flow}

\subsection{Orbiting motion}
\label{app:CircularMotion}

In this section, we restrict ourselves to the Schwarzschild
metric for simplicity.
We want to prescribe a four-velocity for circularly orbiting matter in a thick disk.
A circular four-velocity $u^\mu$ and its associated 1-form $u_\mu$ take the general form
\begin{subequations}
\label{eq:CircularVelocity}
\begin{align}
	u_\mathrm{circ}^\mu\pd_\mu&=u^t\pa{\pd_t+\Omega\pd_\phi},\\
	u^\mathrm{circ}_\mu\ed x^\mu&=-u_t\pa{-\ed t+\ell\ed\phi},
\end{align}
\end{subequations}
where $\Omega=u^\phi/u^t$ denotes the angular velocity and $\ell=-u_\phi/u_t$ the specific angular momentum of the flow.

\subsubsection{Equatorial Keplerian motion}

Timelike circular geodesic motion in the equatorial plane (i.e., Keplerian motion) is well-known \citep{Bardeen1973}.
The four-velocity takes the form \eqref{eq:CircularVelocity} with
\begin{align}
	\label{eq:KeplerianMotion}
	-u_t=\frac{r-2}{\sqrt{r(r-3)}},\qquad
	\ell=\frac{r^{3/2}}{r-2},\qquad
	\Omega=r^{-3/2}.
\end{align}
It is obvious from these relations that Keplerian motion is not allowed inside the photon shell at $r=3$.
Moreover, Keplerian motion is only stable outside the innermost stable circular orbit (ISCO) located at $r_\mathrm{ISCO}=6$.

Below the ISCO, \citet{Cunningham1975} prescribes that the geodesic constants of motion $-u_t$ and $u_\phi$ (the energy and spin angular momentum) keep their ISCO value, which ensures that the flow remains continuous across the ISCO as it spirals into the horizon.
This prescription results in the four-velocity
\begin{align}
	\label{eq:CunninghamPrescription}
	u_\mu^\mathrm{C75}\ed x^\mu=u_t^\mathrm{ISCO}\ed t+u_r\ed r+u_\phi^\mathrm{ISCO}\ed\phi,
\end{align}
where the $(t,\phi)$ components $u_{t,\phi}^\mathrm{ISCO}\equiv u_{t,\phi}(r=r_\mathrm{ISCO})$ take their Keplerian values from Eq.~\eqref{eq:KeplerianMotion} evaluated at the ISCO radius, while the radial component is fixed by unit-normalization:
\begin{align}
	u_r^2=g_{rr}\br{-1-g^{tt}\pa{u_t^\mathrm{ISCO}}^2-g^{\phi\phi}\pa{u_\phi^\mathrm{ISCO}}^2}.
\end{align}
This defines a circular equatorial flow at every radius outside the event horizon $r_\mathrm{H}$: its four-velocity is given by Eqs.~\eqref{eq:CircularVelocity}--\eqref{eq:KeplerianMotion} for $r\geq r_\mathrm{ISCO}$, and by Eq.~\eqref{eq:CunninghamPrescription} for $r_\mathrm{H}<r< r_\mathrm{ISCO}$.

\subsubsection{Keplerian thick disk}

We now wish to find a natural way of extending the equatorial Keplerian disk to a geometrically thick configuration.

For cylindrical radii outside the ISCO ($\rho>6$), we simply define the specific angular momentum $\ell$ to take the same value at every height $z$ above the equator as it does in the equatorial Keplerian disk model.

That is, for $\rho>6$, we impose the axially symmetric profile
\begin{align}
	\ell(\rho,z)=\frac{\rho^{3/2}}{\rho-2},
\end{align}
resulting in a circular four-velocity of the type \eqref{eq:CircularVelocity},
\begin{align}
	\label{eq:ThickDiskOutsideISCO}
	u^\mathrm{thick;>ISCO}_\mu\ed x^\mu=-u_t(\rho,z)\pa{-\ed t+\frac{\rho^{3/2}}{\rho-2}\ed\phi},
\end{align}
whose unit-normalization fixes the time component to be
\begin{align}
	\label{eq:utOutsideISCO}
	-u_t(\rho,z)=\pa{-g^{tt}-\frac{\rho}{(\rho-2)^2}}^{-1/2},
\end{align}
where we used the fact that $g^{\phi\phi}=\pa{r\sin{\theta}}^{-2}=\rho^{-2}$.
Note that $g^{tt}$ depends on $z$ via the spherical radius $r=\sqrt{\rho^2+z^2}$.
It is easy to check that this quantity is well-defined for $\rho>6$ (i.e., that the expression in parentheses remains positive).

For cylindrical radii inside the ISCO ($\rho<6$), a natural extension of the \citet{Cunningham1975} prescription is to require that, at any given height $z$, the components $u_t$ and $u_\phi$ keep their ISCO values for all radii $\rho<6$.
That is,
\begin{align}
	u^\mathrm{thick;<ISCO}_\mu\ed x^\mu=u_t^\mathrm{ISCO}(z)\ed t+u_r(\rho,z)\ed r+u_\phi^\mathrm{ISCO}(z)\ed\phi,
\end{align}
where the $(t,\phi)$ components $u_{t,\phi}^\mathrm{ISCO}(z)\equiv u_{t,\phi}(\rho=\rho_\mathrm{ISCO},z)$ are obtained by evaluating Eqs.~\eqref{eq:ThickDiskOutsideISCO}--\eqref{eq:utOutsideISCO} at the ISCO radius $\rho_\mathrm{ISCO}=6$, while the radial component is again fixed by unit-normalization:
\begin{align}
	\br{u_r(\rho,z)}^2=g_{rr}\pa{-1-g^{tt}\br{u_t^\mathrm{ISCO}(z)}^2-g^{\phi\phi}\br{u_\phi^\mathrm{ISCO}(z)}^2}.
\end{align}
A simple numerical investigation of this relation reveals that it quickly turns negative as the height $z$ above the disk increases, so that this four-velocity is not well-defined.
We must therefore find a new prescription that produces a well-defined flow everywhere.

\subsubsection{General thick disk}

We will start from a general circular four-velocity of type \eqref{eq:CircularVelocity},
\begin{align}
	u^\mathrm{circ}_\mu(\rho,z)\ed x^\mu=-u_t(\rho,z)\pa{-\ed t+\ell(\rho)\ed\phi},
\end{align}
which assumes that the specific angular momentum depends only on cylindrical radius.
The function $\ell(\rho)$ is undetermined at this stage, but it is constrained by the requirement that this four-velocity remain well-defined everywhere outside the horizon.

Its angular velocity is
\begin{align}
	\label{eq:AngularVelocity}
	\Omega\equiv\frac{u^\phi}{u^t}
	=\frac{g^{\phi\phi}u_\phi}{g^{tt}u_t}
	=\frac{\ell}{\rho^2}\pa{1-\frac{2}{r}},
\end{align}
and unit-normalization fixes its time component to be
\begin{align}
	-u_t=\pa{-g^{tt}-g^{\phi\phi}\ell^2}^{-1/2}
	=\sqrt{\frac{1-2/r}{1-\Omega\ell}},
\end{align}
which is well-defined outside the horizon ($r>2$) provided that
\begin{align}
	\Omega\ell<1,
\end{align}
or equivalently [by Eq.~\eqref{eq:AngularVelocity}],
\begin{align}
	\label{eq:RealityCondition}
	\frac{\ell^2}{\rho^2}\pa{1-\frac{2}{r}}<1.
\end{align}
Since $\rho=r\sin{\theta}$, we have $\rho-2\sin{\theta}>0$ outside the horizon, where the condition \eqref{eq:RealityCondition} becomes
\begin{align}
	\label{eq:Condition}
	\ab{\ell(\rho)}<\frac{\rho^{3/2}}{\sqrt{\rho-2\sin{\theta}}}.
\end{align}

We have thus far imposed no conditions on the function $\ell(\rho)$.
At this point, we will assume that it converges to the Newtonian profile $\ell_\mathrm{Newton}=\rho^{1/2}$ at spatial infinity.
We also want to consider a family of profiles that (i) includes the Keplerian profile \eqref{eq:KeplerianMotion} as a special case and (ii) such that the constraint \eqref{eq:Condition} leads to a simple condition.
We are thus naturally led to consider the ansatz
\begin{align}
	\label{eq:lalpha}
	\ell_\alpha(\rho)=\frac{\rho^{3/2}}{\rho+\alpha},
\end{align}
where $\alpha$ is a real constant such that $\alpha=2$ recovers the Keplerian profile \eqref{eq:KeplerianMotion}.
However, we want our profile to be defined at every spacetime point and to remain free of singularities.
We will thus require that $\alpha>0$, so that the denominator may never vanish.
This also ensures that $\ell>0$, and reduces condition \eqref{eq:Condition} to
\begin{align}
	\pa{\rho+\alpha}^2-\rho+2\sin{\theta}>0.
\end{align}
Demanding that the discriminant of this quadratic equation be strictly negative results us in a stronger condition
\begin{align}
	\alpha>\frac{1}{4},
\end{align}
which is the final condition for our angular momentum profile to be well-defined everywhere outside the horizon.
Following \citet{Gold2020}, we consider $\alpha=1$ in this article.
We have numerically checked that this choice leads to a well-defined four-velocity even when the spin is nonzero.

Finally, we note that there is still a locus of spacetime where our chosen profile \eqref{eq:lalpha} produces singular behavior: the axis $\rho=0$.
Taking the limit $\rho\to0$ keeping $z>2$ fixed results in
\begin{align}
	u^t\approx\pa{1-\frac{2}{r}}^{-1/2},\qquad
	u^\phi\approx\rho^{-1/2}\pa{1-\frac{2}{r}}^{1/2},
\end{align}
so the $\phi$ component of the four-velocity, as well as the angular velocity $\Omega$, diverge in the limit $\rho\to0$.
Indeed, Eq.~\eqref{eq:AngularVelocity} shows that the angular velocity grows like $\Omega\approx\ell\rho^{-2}\approx\rho^{-1/2}$ near the axis for the Keplerian $\ell$, which behaves as $\ell\approx\rho^{3/2}$.
Hence, any family $\ell(\rho)$ that includes the Keplerian profile as a member must have a divergent $\Omega$ near $\rho=0$.
However, since the orbiting models that we consider in this paper always have vanishingly small emission close to the axis, this pathology never causes problems.

\subsection{Infalling geodesic motion}

In this section, we work in the Kerr metric and consider
\begin{subequations}
\begin{gather}
	u_\mathrm{rad}^\mu\pd_\mu=u^t\pd_t+u^r\pd_r+u^\phi\pd_\phi,\\
	u_t=-1,\qquad
	u_\phi=0.
\end{gather}
\end{subequations}
This four-velocity describes particles that fall into the black hole from spatial infinity, where they start out with vanishing velocity and angular momentum.
The $(t,\phi)$ contravariant components are
\begin{align}
	u^t=g^{tt}u_t
	=-g^{tt},\qquad
	u^\phi=g^{t\phi}u_t
	=-g^{t\phi}.
\end{align}
Note that $u^\phi$ must be nonzero when the black hole is rotating (frame-dragging prevents purely radial geodesic infall).
Unit-normalization fixes the $r$ covariant component to be
\begin{align}
	u_r=-\sqrt{\frac{-1-g^{tt}}{g^{rr}}},
\end{align}
where we picked the negative sign of the square root to describe infall, and then the $r$ contravariant component is
\begin{align}
	u^r=g^{rr}u_r
	=-\sqrt{\pa{-1-g^{tt}}g^{rr}}.
\end{align}

These formulas are easier to interpret in the Schwarzschild case, where they simplify to
\begin{align}
	u^t=\pa{1-\frac{2}{r}}^{-1},\qquad
	u^r=-\sqrt{\frac{2}{r}},\qquad
	u^\phi=0.
\end{align}
This results in a purely radial velocity (since $u_\phi=0$ at zero spin)
\begin{align}
	\mathbf{v}=v^r\mathbf{e}_r
	=\frac{\ed r}{\ed t}\mathbf{e}_r
	=\frac{u^r}{u^t}\mathbf{e}_r,
\end{align}
where $\mathbf{e}_r$ the unit spacelike vector $\mathbf{e}_r=\pd_r/\sqrt{g_{rr}}$ and
\begin{align}
	v^r=-\sqrt{\frac{2}{r}}\pa{1-\frac{2}{r}}
	\stackrel{r\to\infty}{\approx}\sqrt{\frac{2}{r}},
\end{align}
which at spatial infinity reduces to the Newtonian formula for velocity derived from classical conservation of energy:
\begin{align}
	\frac{1}{2}\pa{v^r}^2-\frac{1}{r}=0.
\end{align}

\section{Precision of ray tracing computations}
\label{app:gyoto_prec}

\begin{figure}
	\centering
	\includegraphics[width=\columnwidth]{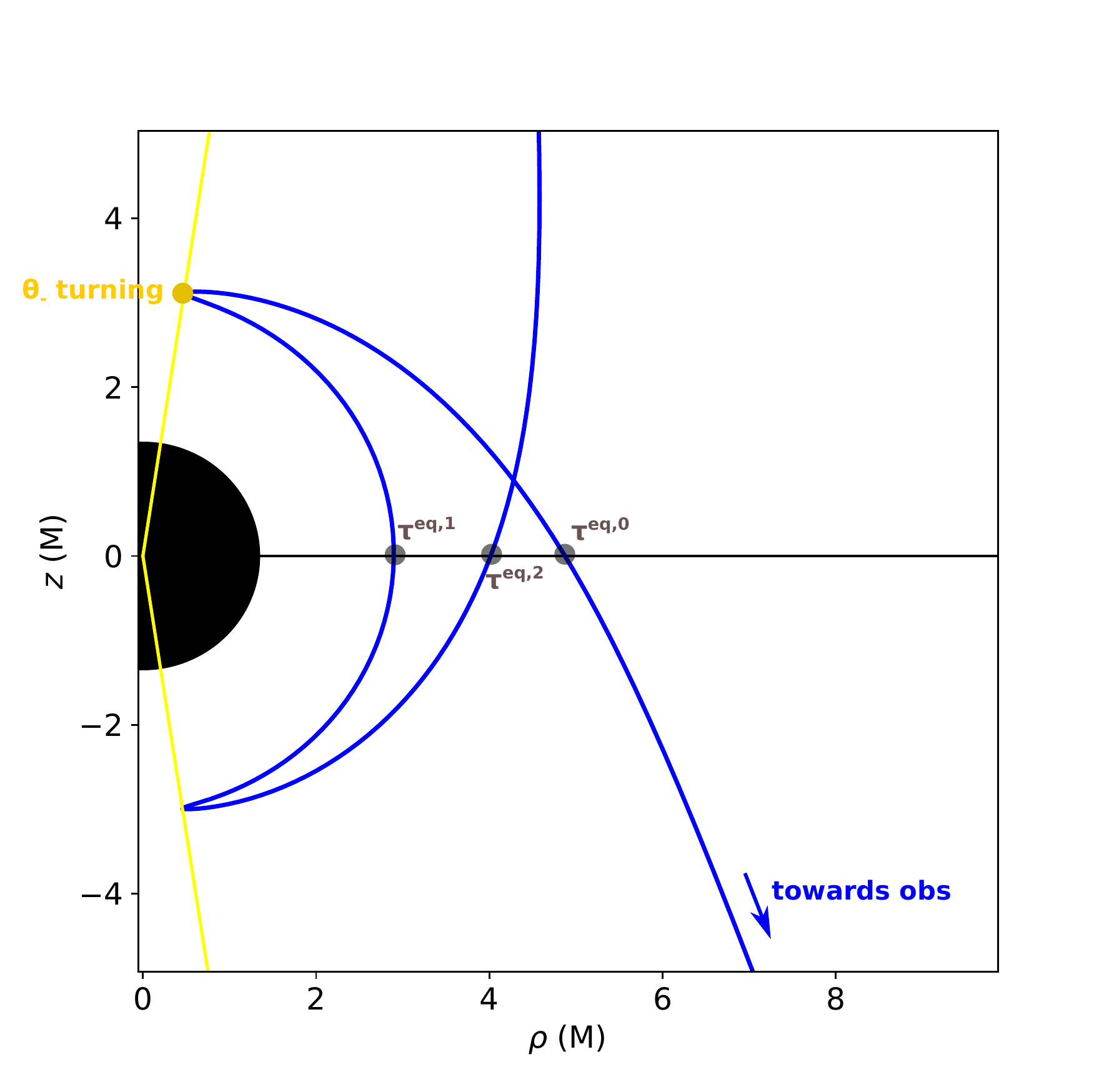} 
	\caption{A null geodesic of order $n=2$ (two $\theta$ turning points), in blue.
	The black disk is the event horizon and the yellow lines indicate the angular turning points $\theta_-$ and $\theta_+=\pi-\theta_-$, which are computed from the analytical expression for $\theta_-$ as a function of black hole spin and the geodesic constants of motion~\citep[see, e.g.,][]{GrallaLupsasca2019a}.
	The three equatorial crossings of the geodesic are labeled by the black dots.} 
	\label{fig:HighOrderGeodesics}
\end{figure}

In this section, we examine the strongly lensed null geodesics in Kerr, i.e., those that explore the shell of spherical photon orbits before escaping to produce high-order images.
Their trajectories are very highly bent, as illustrated in Fig.~\ref{fig:HighOrderGeodesics}.

Kerr null geodesics obey \citep[e.g.,][]{GrallaLupsasca2019a,GrallaLupsasca2019b}
\begin{align}
	\label{eq:NullGeodesics}
	I_r\equiv\fint_{r_s}^{r_o}\frac{\ed r}{\pm_r\sqrt{\mathcal{R}(r)}}
	=\fint_{\theta_s}^{\theta_o}\frac{\ed\theta}{\pm_\theta\sqrt{\Theta(\theta)}}
	\equiv G_\theta,
\end{align}
where $\mathcal{R}(r)$ and $\Theta(\theta)$ are radial and angular geodesic potentials, and the integrals are evaluated along the entire trajectory with the signs $\pm_{r,\theta}$ flipping at turning points of the radial/angular motion, which correspond to zeros of their respective potentials.

\begin{figure}
	\centering
	\includegraphics[width=\columnwidth]{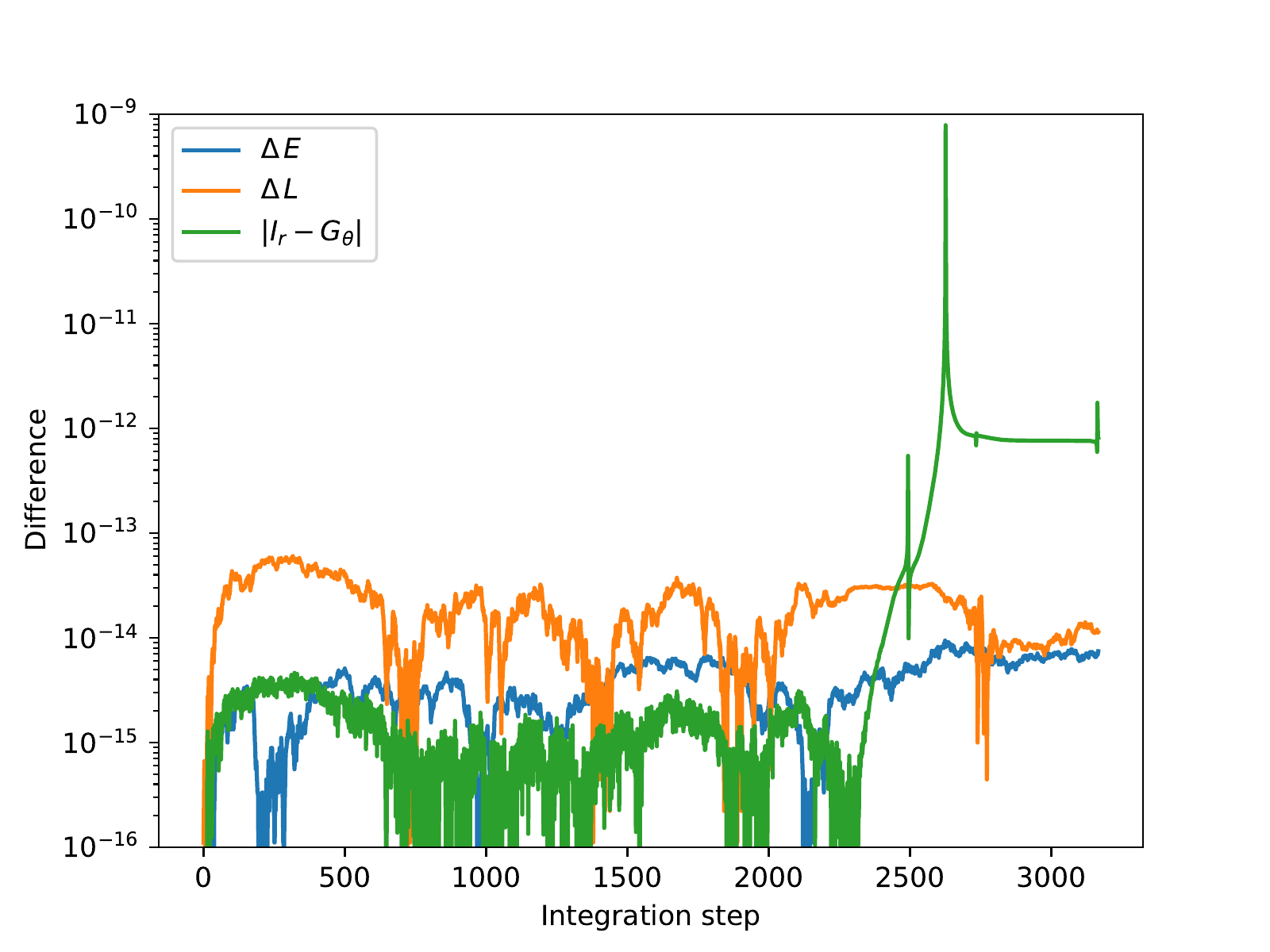} 
	\caption{Evolution of the difference between the initial and current values of the geodesic constants $E=-p_t$ (blue) and $L=p_\phi$ (orange), as well as the quantity $|I_r - G_\theta|$ (green), as a function of the integration step.
	The peak of $|I_r - G_\theta|$ happens close to an angular turning point where the geodesic evolves in the Kerr spherical null geodesics region.} 
	\label{fig:GeodesicError}
\end{figure}

We consider a geodesic traveling from a source at $(r_s,\theta_s)$ to an observer at $(r_o,\theta_o)$.
To assess the precision of our geodesic integrator, we plot in Fig.~\ref{fig:GeodesicError} the numerical evolution along the geodesic of the quantity $|I_r-G_\theta|$, as well as of the conserved energy $E=-p_t$ and azimuthal angular momentum $L=p_\phi$.

These quantities should remain constant along the geodesic, and the figure shows that their error is consistently below $10^{-13}$, except for a peak of $10^{-9}$ in $|I_r-G_\theta|$ close to an angular turning point.
We have also checked that the radii of the three equatorial crossings match the analytical formula derived by \citet{GrallaLupsasca2019b}, which gives the radius of the equatorial crossing after $m$ angular turns in terms of the Jacobi elliptic sine $\sn(x|k)$:
\begin{align}
	r_\mathrm{eq}^{(m)}=\frac{r_4r_{31}-r_3r_{41}\sn^2\pa{\left.\frac{1}{2}\sqrt{r_{31}r_{42}}\tau_\mathrm{eq}^{(m)}-\mathcal{F}_o\right|k}}{r_{31}-r_{41}\sn^2\pa{\left.\frac{1}{2}\sqrt{r_{31}r_{42}}\tau_\mathrm{eq}^{(m)}-\mathcal{F}_o\right|k}}.
\end{align}
Here, $r_{ij}=r_i-r_j$ with $\{r_1,r_2,r_3,r_4\}$ the four roots of the quartic potential $\mathcal{R}(r)$ \citep[with exact expressions given in Eq.~(A8) of][]{GrallaLupsasca2019a}, and $\mathcal{F}_o$ is the elliptic integral of the first kind
\begin{align}
	\mathcal{F}_o=F\pa{\left.\arcsin\sqrt{\frac{r_{31}}{r_{41}}}\right|k},\qquad
	k=\frac{r_{32}r_{41}}{r_{31}r_{42}}.
\end{align}
The Mino time elapsed up to the $m^\text{th}$ equatorial crossing is
\begin{align}
	\tau_\mathrm{eq}^{(m)}=\frac{2mK-\sign\pa{p_o^\theta}F_o}{a\sqrt{-u_-}},
\end{align}
where $p_o^\theta$ is the polar photon momentum at the observer,
\begin{align}
	u_\pm=\Delta_\theta\pm\sqrt{\Delta_\theta^2+\frac{\eta}{a^2}},\qquad
	\Delta_\theta=\frac{1}{2}\pa{1-\frac{\eta+\lambda^2}{a^2}},
\end{align}
are the zeros $u=\cos^2{\theta}$ of $\Theta(u)$ in terms of the energy-rescaled angular momentum $\lambda=L/E$ and Carter constant $\eta=Q/E^2$, and
\begin{align}
	F_o=F\pa{\left.\arcsin\pa{\frac{\cos{\theta_o}}{\sqrt{u_+}}}\right|\frac{u_+}{u_-}},\qquad
	K=F\pa{\left.\frac{\pi}{2}\right|\frac{u_+}{u_-}}
\end{align}
are an elliptic integral of the first kind and its completion $K$.

We found that the three numerical and analytical values of $r_\mathrm{eq}^{(m)}$ agree to within $10^{-4}-10^{-5}M$.
Note that the equatorial crossing labeled $\tau^{\mathrm{eq},2}$ in Fig.~\ref{fig:HighOrderGeodesics} occurs after the backward ray traced geodesic has encountered two angular turning points; this shows that a very high level of numerical precision is maintained despite the spike in the error $|I_r-G_\theta|$ at that crossing.
Finally, we have compared the value of the radial turning point $r_4$ (i.e., the minimum value of the radial trajectory) to the analytical expression given in Eq.~(A8d) of \citet{GrallaLupsasca2019a}, and found an error $\lesssim10^{-8}M$.
All these results lead us to conclude that our null geodesic integrator is highly accurate, even within the photon shell of bound spherical photon orbits.

Besides the geodesic integration, \textsc{Gyoto} also integrates the radiative transfer equation by discretizing the part of the geodesic that intersects the disk, using a constant step of size $\delta=0.1M$.
We have checked that reducing this step size by a factor of $10$ changes the observed specific intensity by less than 0.1\%, which is sufficient given that our synchrotron emissivity prescription is only at the percent level of precision (see App.~\ref{app:Synchrotron}).

\section{Image orders}
\label{app:ImageOrder}

Null geodesics in Kerr can be conveniently parametrized by the so-called Mino time \citep{GrallaLupsasca2019a,GrallaLupsasca2019b}
\begin{align}
	\tau=I_r
	=G_\theta,
\end{align}
with $I_r$ and $G_\theta$ are defined in Eq.~\eqref{eq:NullGeodesics}.
Before the first angular turning point (i.e., between the points $S$ and $T_1$ in Fig.~\ref{fig:DiskOrders}), the Mino time elapsed along the geodesic as it travels backwards from the observer at $\theta_o$ to $\theta=\theta_s$ is \citep{GrallaLupsasca2019b}
\begin{align}
	\tau(\theta)=\sign\pa{p_s^\theta}\pa{\mathcal{G}_\theta^o-\mathcal{G}_\theta^s},
\end{align}
where $p_s^\theta$ is the polar photon momentum at $\theta=\theta_s$, while
\begin{align}
	\mathcal{G}_\theta=-\frac{1}{a\sqrt{-u_-}}F\pa{\left.\arcsin\pa{\frac{\cos{\theta}}{\sqrt{u_+}}}\right|\frac{u_+}{u_-}},
\end{align}
and $\mathcal{G}_\theta^o$ and $\mathcal{G}_\theta^s$ are evaluated at the $\mathcal{G}_\theta$ evaluated at $\theta=\theta_o$ and $\theta=\theta_s$, respectively.
We can determine the Mino time $\tau_1$ elapsed at the first angular turning point by setting
\begin{align}
	\theta_s=\arccos\pa{\pm\sqrt{u_+}},\qquad
	\pm=\sign\pa{p_o^\theta},
\end{align}
while the Mino time elapsed between two successive angular is
\begin{align}
	\Delta\tau=\frac{2K}{a\sqrt{-u_-}}.
\end{align}

The order of any null geodesic is then defined as
\begin{itemize}
	\item[\textbullet] $n=0$: between the observer and $\tau=\tau_1$,
	\item[\textbullet] $n=1$: between $\tau=\tau_1$ and $\tau=\tau_1+\Delta\tau$,
	\item[\textbullet] $n=2$: between $\tau=\tau_1+\Delta\tau$ and $\tau=\tau_1+2\Delta\tau$.
\end{itemize}
The integration is cut off past $n=2$, so as not to pollute the image by unresolved $n>2$ features.

We are thus able to provide not only the full image (containing all layers $n\leq2$), but also individual images of each layer $n\in\{0,1,2\}$ consisting of photons loaded onto each null geodesic at the corresponding order.

\end{document}